\documentclass[fleqn,usenatbib]{mnras}

\usepackage{newtxtext,newtxmath}
\usepackage{subfigure}
\usepackage[T1]{fontenc}
\usepackage{ae,aecompl}

\usepackage{graphicx}
\usepackage{amsmath}
\usepackage{amssymb}
\usepackage{caption}
\usepackage{soul}
\usepackage[colorinlistoftodos]{todonotes}

\title[Size growth of galaxies hosting the first quasars]{How AGN feedback drives the size growth of the first quasars}
\author[D. van der Vlugt \& T. Costa]{
Dieuwertje van der Vlugt$^1$\thanks{E-mail:\, dvdvlugt@strw.leidenuniv.nl} \& Tiago Costa$^{2,1}$
\\
$^1$Leiden Observatory, Leiden University, P.O. Box 9513, 2300 RA Leiden, the Netherlands\\
$^2$ Max-Planck-Institut f\"ur Astrophysik, Karl-Schwarzschild-Stra{\ss}e 1, D-85748 Garching b. M\"unchen, Germany
}
\pubyear{2019}

\begin{document}
\label{firstpage}
\pagerange{\pageref{firstpage}--\pageref{lastpage}}
\maketitle
\begin{abstract}
Quasars at $z \,=\, 6$ are powered by accretion onto supermassive black holes with masses $M_{\rm BH} \sim 10^9 \rm \, M_{\odot}$. Their rapid assembly requires efficient gas inflow into the galactic nucleus, sustaining black hole accretion at a rate close to the Eddington limit, but also high central star formation rates. Using a set of cosmological `zoom-in' hydrodynamic simulations performed with the moving mesh code {\sc Arepo}, we show that $z \,=\, 6$ quasar host galaxies develop extremely tightly bound stellar bulges with peak circular velocities $300 \-- 500 \, \rm km \, s^{-1}$ and half-mass radii $\approx 0.5 \, \rm kpc$. Despite their high binding energy, we find that these compact bulges expand at $z \, < \, 6$, with their half-mass radii reaching $ \approx 5 \, \rm kpc$ by $z \, = \, 3$. The circular velocity drops by factors $\approx 2$ from their initial values to $200 \-- 300 \, \rm km \, s^{-1}$ at $z \, \approx \, 3$ and the stellar profile undergoes a cusp-core transformation. By tracking individual stellar populations, we find that the gradual expansion of the stellar component is mainly driven by fluctuations in the gravitational potential induced by bursty AGN feedback. We also find that galaxy size growth and the development of a cored stellar profile does not occur if AGN feedback is ineffective.
Our findings suggest that AGN-driven outflows may have profound implications for the internal structure of massive galaxies, possibly accounting for their size growth, the formation of cored ellipticals as well as for the saturation of the $M_{\rm BH} \-- \sigma_{\rm \star}$ seen at high velocity dispersions $\sigma_{\rm \star}$.
\end{abstract}

\section{Introduction}
\label{sec:Introduction}

Detections of bright quasars at $z \, > \, 6$ show that supermassive black holes with masses $M_{\rm BH} \, \gtrsim \, 10^9 \, \rm M_\odot$ assemble in less than a Gyr after the Big Bang \citep[e.g.][]{Fan_2004, Willott_2010, Mortlock_2011, Venemans_2013, Gallerani_2017}. The most distant quasar known has $z \, = \, 7.5$ \citep{Banados_2018}, while the most massive black hole at $z > 6$ has $M_{\rm BH} \gtrsim 10^{10} \, \rm M_\odot$ \citep{Wu_2015}.
The extreme conditions which are likely required to account for such rapid black hole growth provide interesting testbeds for models of black hole accretion and its associated active-galactic-nucleus (AGN) feedback.

If the main growth channel is accretion of interstellar gas, the black hole accretion rate must be close to the Eddington limit for a significant fraction of the Hubble time at $z > 6$ even if black hole seeds are massive \citep[e.g][]{Sijacki_2009}.
For lower mass seed black holes, prolonged episodes of super-Eddington accretion rates are likely required to achieve rapid enough black hole growth.
Such constraints require supermassive black holes to reside in galaxies where gas in-fall towards their nuclei is sustained, a condition which is most easily satisfied if their host dark matter haloes are massive with $M_{\rm DM} \, \gtrsim \, 10^{12} \, \rm M_\odot$ \citep{Efstathiou_1988, Volonteri_2006, Costa_2014}.

Cosmological hydrodynamic simulations following black hole growth as well as the impact of energy deposition by the quasar (AGN feedback) have succeeded in forming massive black holes with $M_{\rm BH} \, \gtrsim \, 10^9 \, \rm M_\odot$ by $z \, = \, 6$ \citep{Sijacki_2009, DiMatteo_2012, Costa_2014}.
The efficient in-fall of gas into the galactic nucleus necessary to sustain black hole growth, however, also likely triggers powerful central starbursts and the formation of compact stellar bulges characterised by extreme stellar velocity dispersions $\sigma_{\rm *}$.
In their high resolution cosmological simulations, \citet{Dubois_2012} find galaxies with stellar bulges with $\sigma_{\rm *} > 600 \, \rm km \, s^{-1}$ measured within $\approx 500 \, \rm pc$ embedded within haloes with virial mass $M_{\rm vir} \, \gtrsim \, 10^{12} \, \rm M_\odot$ at $z > 6$. The compact sizes are attributed to their special location at the intersection between multiple large-scale cold gas filaments, a configuration which leads to efficient angular momentum cancellation.

Extremely compact galaxies appear to be the general outcome of galaxy formation in massive haloes at $z \,\gtrsim\, 6$. \citet{Costa_2015} show, using similar simulations, that 'compact bulges' with peak circular velocities $\gtrsim 500 \, \rm km \, s^{-1}$ form even in the presence of supernova and AGN feedback.
Using $\sim 10 \, \rm pc$ resolution simulations, \citet{Curtis_2016} find peak circular velocities as high as $\approx 900 \, \rm km \, s^{-1}$ for the 'compact bulge' at $z \,\approx\,5$, while \citet{DiMatteo_2017} show that extremely compact galaxies assemble already by $z \, = \, 8$.

Based on the extreme binding energy of these systems, we are led to question whether they may be able to survive down to lower redshift. 
Using the {\sc Illustris} simulations, \citet{Wellons_2016} identify various evolutionary paths for high-z massive, compact galaxies.
These include (i) the survival of the core within a subsequently accreted stellar envelope, (ii) its destruction in a major merger, (iii) its consumption by a more massive galaxy and (iv) its undisturbed survival down to $z \, = \, 0$.
In addition, \citet{Wellons_2016} find that the compact galaxies grow both in mass and size, which they attribute to accretion of ex-situ stars, a process identified in earlier work \citep{Naab_2009, Oser_2010, Johansson_2012}.
According to these studies, most descendants of high-z massive, compact galaxies are no longer compact at $z \, = \, 0$.
Whether and how this picture may change for the extreme systems which form in high-$\sigma$ peaks at $z > 6$, however, remains unexplored. 

Previous studies have proposed quasar feedback itself as a mechanism for galactic growth \citep[e.g.][]{Fan_2008}. In this scenario, successive episodes of rapid ejection of gas from a massive galaxy lead to gravitational potential fluctuations and the expansion of the collisionless component (see also \citealt{Navarro_1996, RagoneFigueroa_2011, Teyssier_2013, Pontzen_2014, Ogiya_2014, Penoyre_2017}). Others have proposed `positive feedback' as a possible channel for massive galaxy growth \citep{Ishibashi_2013}.

In this paper, we investigate the fate of the compact galaxies hosting bright $z > 6$ quasars down to $z \, = \, 3.3$, using a suite of zoom-in simulations targeting massive $M_{\rm vir} \gtrsim 3 \times 10^{12} \, \rm M_\odot$ haloes at $z \, = \, 6$.
We start by describing our suite of simulations in Section~\ref{sec:Simulations} and then present our main findings in Section~\ref{sec:Results}.
The implications of our results are discussed in Section~\ref{sec:Discussion} and our main conclusions are summarised in Section~\ref{sec:Conclusions}.

\section{Simulations}
\label{sec:Simulations}
We perform cosmological, `zoom-in', hydrodynamic simulations targeting the five most massive haloes found in the dark matter-only Millennium simulation \citep{Springel_2005} at $z = 6.2$ and follow their evolution down to $z \, = \, 3$.
These haloes represent high-$\sigma$ fluctuations of the cosmic density field (see Table \ref{tab:main properties haloes} for a list of their main properties) and are massive enough to ensure rapid black hole growth \citep{Costa_2014}.
We assume a $\Lambda$CDM cosmology with parameters $h=0.73$, $\Omega_m =0.25$, $\Omega_{\Lambda}=0.75$, $\Omega_b =0.041$, identical to the parent Millennium simulation, but a lower $\sigma_8$ value of $0.8$, in better accordance with the Planck cosmology (\citealt{PlanckCollaboration2014}).

\subsection{Numerical Setup}
\label{sec:numerical setup}

The numerical setup of our simulations is identical to that of \citet{Costa_2015} and is here only briefly summarised.

Our cosmological simulations are performed with the moving-mesh code {\sc Arepo} \citep{Springel_2010} and consist of `zoom-ins' of the most massive haloes found in the Millennium volume \citep{Springel_2005} at $z \, = \, 6$.

Besides the collisionless dynamics of dark matter and stellar populations, our simulations follow the hydrodynamic evolution of gas on an unstructured Voronoi mesh using a second-order finite volume scheme. Explicit refinement and de-refinement is employed in order to maintain the masses of gas cells within a factor of two of a target mass, here chosen to be $m_{\rm gas} \, = \, 1.8 \times 10^6 \, h^{-1}\rm M_\odot$ (see Table~\ref{tab:resolution} for a list of the main numerical parameters of our simulations). The geometry and size of the fluid patches is adapted to higher density regions thereby concentrating resolution elements in the densest regions.
For example, in Halo1 cells can be as small as 22 pc at $z \, = \, 6.2$ and $\approx$ 1.5\% of the cells within the virial radius of Halo1 can be classified as `small cells' ($\leq 2 \times \Delta x$, see Table~\ref{tab:resolution}).
Individual high-resolution dark matter particles have a mass $m_{\rm dm} \, = \, 6.8 \times 10^6 h^{-1} \, \rm M_\odot$ and the stellar particles have $m_{\star} \, = \, 1.3 \times 10^5 h^{-1} \, \rm M_\odot$.
We use a fixed comoving gravitational softening $\epsilon_{\rm soft} \, = \, 1 h^{-1} \, \rm ckpc$, which means their physical gravitational softening increases with time.
We also perform a simulation, in which we maintain a constant physical gravitational softening below $z \, = \, 6$.
For the gas component, the gravitational softening length scales with the gas cell size, but is never allowed to be smaller than the collisionless gravitational softening.

We follow radiative cooling for a hydrogen and helium gas in collisional ionization equilibrium as in \citet{Katz_1996}, subjected to a spatially constant, but time-dependent UV background with reionization occurring at $z \approx 6$ (\citealt{Faucher-Giguere_2009}).
Metal-line cooling is treated as in \citet{Vogelsberger_2013} with the minimum temperature down to which gas is allowed to cool radiatively set to $10^3$ K.

We follow \citet{Springel_2003} to model star formation and associated feedback.
Stellar particles are spawned stochastically out of gas particles with number density exceeding a threshold $n_0 \, = \, 0.13 \, \rm cm^{-3}$ and with temperatures $T < 10^5 \, \rm K$.
Star-forming gas follows an effective equation of state resulting from the balance of radiative cooling and cold cloud evaporation within the unresolved interstellar medium.

\begin{table}
\centering
\caption{Main numerical parameters in our simulations. The first, second and third column give the mass of high-resolution dark matter particles, the target mass of gas cells and the mass of stellar particles, respectively. The gravitational softening length is given in comoving units in the fourth column. In the redshift range investigated here, this corresponds to 150 - 400 physical pc. The minimum cell size is given in the fifth column. 
}
\begin{tabular}{ccccc}
	\hline
	$m_{\rm dm}$ & $m_{\rm gas}$ & $m_{\star}$ &$\epsilon_{\rm soft}$ & $\Delta x$\\
     ($h^{-1}\rm M_{\odot}h$) & ($h^{-1}\rm M_{\odot}$) & ($h^{-1}\rm M_{\odot}$)& ($h^{-1}$ckpc) & (pc) \\
	\hline
	 $6.8 \times 10^6$ & $1.8 \times 10^6$ & $1.3 \times 10^5$ & 1 & 22\\
        \hline
\end{tabular}
\label{tab:resolution}
\end{table}

Supernova-driven winds are modelled following \cite{Springel_2003, Vogelsberger_2013}, i.e. a gas is launched from star forming regions at a rate of $\dot{M} = \eta \dot{M}_{\star}$, where $\dot{M}_{\star}$ is the star formation rate and $\eta$ is the mass-loading factor.
Gas cells ejected from a star-forming region are converted into wind particles and launched isotropically.
We assume a mass loading factor of $\eta = 1$ and kick-velocity of $1200 \, \rm km \, s^{-1}$ (as in \citealt{Costa_2015}).
Wind particles interact with other matter components only gravitationally until they travel to a region where the local density drops below $0.05 n_{\rm 0}$ or a maximum travel time is reached.
When either of these criteria are met, the wind particle's mass, metalicity, momentum and energy are deposited into its host gas cell.

\begin{table*}
\centering
\caption{Main properties of the sample of the 6 haloes followed by our simulations, evaluated at $z=6.2$ and $z=3.3$ in, respectively, the top and bottom table. We list the mass of the most massive black hole present in each volume (second column), the total mass within the virial radius (third column), virial radius (fourth column), the virial temperature (fifth column) of its parent halo, the total star formation rate within the virial radius (sixth column) and the stellar half mass radius as predicted by {\sc SUBFIND} (seventh column).}
\label{tab:main properties haloes}
\begin{tabular}{lccccccr}
\hline
Simulation at $z=6.2$ & $M_{\rm BH}$ & $M_{200}$ & $R_{200}$ &$T_{200}$ & $SFR_{\rm 200}$ & $V_{\rm 200}$ & $R_{\rm eff}$\\
         & ($\times 10^8 \rm M_{\odot}$) & ($\times 10^{12} \rm M_{\odot}$) & (kpc)& ($\times 10^6$ K) & ($\rm M_{\odot} yr^{-1}$) & ($\rm km$ $\rm s^{-1}$) & (kpc) \\
\hline
Halo1 & 8.81 & 3.97 & 70.0 & 4.01 & 352.82 & 494.14 & 1.16\\
Halo1-NoAGN & - & 3.97 & 70.0 & 4.24 & 396.59 & 494.02 & 1.05\\
Halo2 & 9.12 & 1.81 & 59.85 & 4.03 & 215.0 & 422.61 & 0.80\\
        Halo3 & 17.8 & 3.73 & 68.54 & 3.25 & 237.56 & 483.92 & 0.59\\
        Halo4 & 7.43 & 3.18 & 64.96 & 3.24 & 520.49 & 458.68 & 0.91\\
        Halo5 & 9.66 & 1.95 & 55.21 & 3.02 & 70.81 & 389.78 & 0.76\\
        \hline
\end{tabular}
\begin{tabular}{lccccccr}
\hline
Simulation at $z=3.3$  & $M_{\rm BH}$ & $M_{200}$ & $R_{200}$ &$T_{200}$ & $SFR_{\rm 200}$ & $V_{\rm 200}$ & $R_{\rm eff}$\\
         & ($\times 10^8 \rm M_{\odot}$) & ($\times 10^{12} \rm M_{\odot}$) & (kpc) & ($\times 10^6$ K) & ($\rm M_{\odot} yr^{-1}$) & ($\rm km$ $\rm s^{-1}$) & (kpc) \\
\hline
Halo1 & 85.0 & 31.2 & 230.03 & 8.71 & 1434.12 & 764.26 & 6.16\\
Halo1-NoAGN & - & 31.7 & 231.08 & 7.07 & 2418.61 & 767.77 & 0.64\\
Halo2 & 80.8 & 20.8 & 200.92 & 9.72 & 93.38 & 667.10 & 6.73\\
        Halo3 & 74.4 & 20.4 & 199.61 & 7.52 & 199.35 & 662.76 & 5.52\\
        Halo4 & 86.1 & 21.1 & 201.81 & 9.20 & 238.41 & 670.08 & 7.63\\
        Halo5 & 69.3 & 22.1 & 205.01 & 8.84 & 194.9 & 680.70 & 7.98\\
        \hline
\end{tabular}
\end{table*}

Black holes are modelled as sink particles. Black hole accretion occurs at scales much smaller than cosmological simulations typically can resolve. Resolving the radius at which accretion takes place requires pc resolution which is $\sim \, 2$ orders of magnitude lower than the resolution of our simulations. This means we cannot resolve transfer of angular momentum, gas collapse or even the structure of infalling material. The unresolved physics of accretion is therefore also modelled in a subgrid fashion following \citet{Springel_2005} and \citet{DiMatteo_2005}. The accretion is assumed to be of a Bondi-Hoyle-Lyttleton type. 
Assuming this type of prescription has the advantage that it is a relatively simple parametrisation of black hole accretion. In addition, this type of accretion is commonly assumed in almost every state-of-the-art cosmological simulation (e.g., in {\sc EAGLE} \citep{Schaye_2015}, {\sc HORIZON-AGN} \citep{Dubois_2014},{\sc ILLUSTRIS-TNG} \citep{Weinberger_2017}, {\sc MASSIVE-BLACK} \citep{DiMatteo_2012}, {\sc BLUE-TIDES} \citep{Feng_2016}) with the exception of the {\sc SIMBA} simulations \citep{Dave_2019}. This allows us to connect our results with other studies.

The accretion rate is capped at the Eddington rate for each black hole. The accretion rate is given by
\begin{equation}
\dot{M}_{\rm BH} \, = \, \min {\left[\frac{4\pi \alpha G^2 M^2_{\rm BH} \rho_{\rm gas}}{(c_{\rm s}^2+v^2)^{3/2}}, \dot{M}_{\rm edd}\right]} \, ,
\label{eq:bondi}
\end{equation}
where $\alpha$ is a dimensionless parameter, set to a value of 100, that is introduced to recover a volume coverage of the Bondi rates for the cold and hot interstellar medium phases, $\rho_{\rm gas}$ and $c_{\rm s}$ are the density and sound speed of the gas surrounding the black hole, respectively, $v$ is the speed of the black hole particle relative to the local gas speed and $G$ is the gravitational constant. The sound speed and density are estimated locally at the position of the black hole particle by taking an SPH average over its 64 nearest gas cell neighbours.

The Eddington accretion rate used to find the accretion rate is defined as
\begin{equation}
\dot{M}_{\rm edd} \, = \, \frac{4\pi G M_{\rm BH} m_{\rm p}}{\epsilon_{\rm r} \sigma_{\rm T}c} \, ,
\label{eq:bondi_2}
\end{equation}
where $m_{\rm p}$ is the proton mass, $\sigma_{\rm T}$ the cross-section for Thomson scattering, $c$ the speed of light and $\epsilon_{\rm r}$ the radiative efficiency. We set $\epsilon_{\rm r}$ to the standard value of 0.1

We seed haloes with a virial mass $M_{200} = 10^{10} h^{-1} \rm M_{\odot}$ with seed black hole particles with mass $10^5 h^{-1} \rm M_{\odot}$.
Only one such particle is allowed to form per halo.
Besides accretion, every black hole is allowed to grow by merging with other black holes that fall within its smoothing length, evaluated by averaging over the properties of the 64 nearest gas cell neighbours, and have a relative velocity lower or comparable to the local sound speed.

We model AGN feedback at all mass accretion regimes through an isotropic coupling of a small fraction of the AGN bolometric luminosity to the surrounding gas cells as in \citet{DiMatteo_2005, Sijacki_2009} or as in the `quasar mode' feedback used in {\sc ILLUSTRIS} \citep{Vogelsberger_2013} and {\sc ILLUSTRIS-TNG} \citep{Weinberger_2017} simulations.
Thermal energy is injected continuously into the 64 nearest gas cells at a rate given by
\begin{equation}
\dot{E}_{\rm feed} = \epsilon_f L_{\rm bol} = \epsilon_{\rm f} \epsilon_{\rm r} (1 - \epsilon_{\rm r})^{-1} \dot{M}_{\rm BH}c^2 \, ,
\end{equation}
where $\epsilon_f$ is the feedback efficiency, set to 0.05, which reproduces the normalisation of the $M_{\rm BH} - \sigma_{\star}$ relation, the correlation between the black hole mass and the stellar velocity dispersion, in fully cosmological simulations \citep{DiMatteo_2005, Sijacki_2007, DiMatteo_2008}.

\subsection{Simulation sample}

We perform a total of 8 simulations, in which we explore the evolution of the targeted galaxies after their $z \, = \, 6$ quasar phase in different environments. Our fiducial simulations are named Halo1 to Halo5 and these haloes are simulated till $z \, = \, 3.3$.
We also re-run simulations for Halo1 with a modified setup: (i) switching off black hole accretion and AGN feedback at $z \, = \, 5.2$, (ii) excluding black hole growth and AGN feedback altogether and (iii) switching to fixed physical (as opposed to comoving) softening lengths at $z \, = \, 6$. These simulations are named, respectively, Halo1-weakAGN, Halo1-NoAGN and Halo1-Soft. Halo1-NoAGN is also simulated till $z \, = \, 3.3$ whereas Halo1-weakAGN and Halo1-Soft, simulations with finer time-steps, are only run till $z \, = \, 4.5$.

\section{Results}
\label{sec:Results}

Here we present the main results from our simulations and analysis.
We start with a qualitative description of the evolution of the quasar host galaxies in our simulations.

\subsection{Overview}

\begin{figure*}
\includegraphics[width=1.0\textwidth]{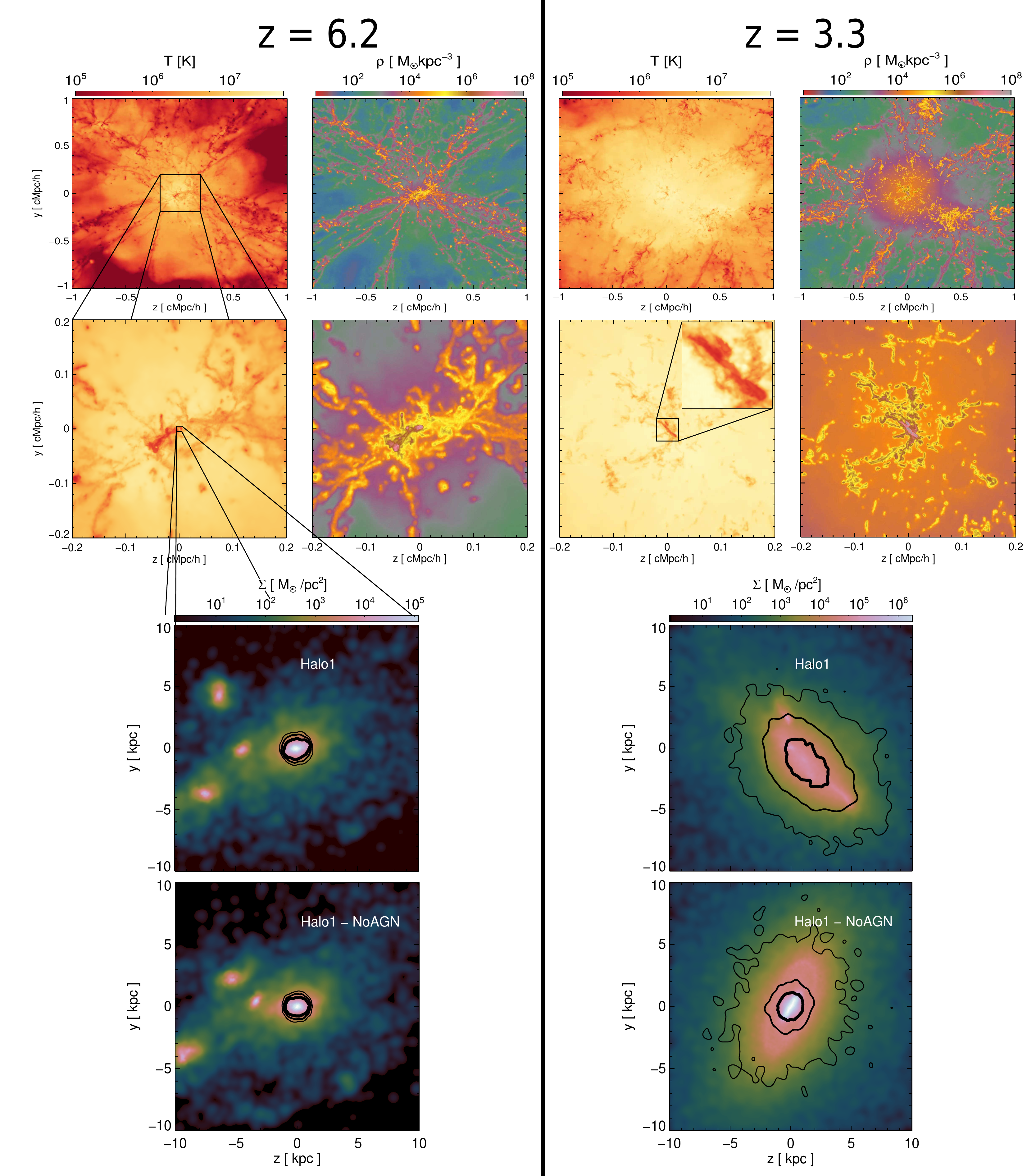}
    \caption{Illustration of the typical cosmological environment of the targeted haloes with AGN feedback in the simulation. The first row shows mass-weighted density and temperature projected along a slab of thickness $2 h^{-1} \, \rm cMpc$. The second row is a zoom-in of the first row projected along a slab of thickness $200 h^{-1} \, \rm ckpc$. The left-hand panels show the environment at $z\,=\, 6.2$ and at the right-hand panels show the environment $z \, = \, 3.3$. The stellar maps of the central galaxy in the simulation with and without AGN feedback at the bottom of the figure are projected along a slab of thickness $100 h^{-1} \, \rm ckpc$, sizes are in physical coordinates. The contours represent the surface density of bulge stars (see text for definition) and correspond to $5, 100, 700 \, \rm M_{\odot} \, pc^{-2}$ surface density levels, where the thickest contour corresponds to the highest surface density.}
    \label{fig:panels}
\end{figure*}

The properties of the galactic haloes targeted by our simulations at $z \, = \, 6$ are summarised in Table \ref{tab:main properties haloes}. The total halo masses are all on the order of a few $10^{12} \, \rm M_\odot$. Their central black holes have grown to masses $5 \times 10^8 \-- 10^9 \rm M_{\odot}$ as has been seen in previous simulations (\citealt{Sijacki_2009, DiMatteo_2012, Costa_2014}). 

By $z \, = \, 3.3$ the total halo masses all grow to $\sim 10^{13} \, \rm M_\odot$, the central black holes masses at this redshift lie in the range $7 \times 10^9 \-- 9 \times 10^9 \rm M_{\odot}$ (see second part of Table \ref{tab:main properties haloes}). The stellar mass within the stellar half mass radius, as predicted by {\sc SUBFIND} (\citealt{Springel_2001}), is $\approx 2.3 \times 10^{10} M_{\odot}$ at $z\,=\, 6.2$ and reaches $\approx 1.5 \times 10^{11} M_{\odot}$ at $z\,=\, 3.3$. 
For comparison, in the simulation without AGN feedback, the targeted galaxy's stellar mass grows from a comparable $\approx 2.2 \times 10^{10} M_{\odot}$ at $z\,=\, 6.2$ to a significantly higher value of $\approx 6.2 \times 10^{11} M_{\odot}$ at $z\,=\, 3.3$.

The typical cosmological environment of the targeted haloes is illustrated in Fig.~\ref{fig:panels}, where we show the mass-weighted density and temperature projected along a slab of thickness $2 h^{-1} \, \rm cMpc$ and $200 h^{-1} \, \rm ckpc$, in the top and the second row respectively, at $z \,=\, 6.2$ (left-hand panels) and at $z \, = \, 3.3$ (right-hand panels).
The massive galaxy resides at the intersection of a number of filaments which feed it with cold gas, as seen in previous simulations (\citealt{Sijacki_2009, DiMatteo_2012, Dubois_2012, Costa_2014}).
Remarkably, the orientation of the filaments changes little between $z \, = \, 6.2$ and $z \, = \, 3.3$, though it is clear that a diffuse component becomes prominent at $z \,=\, 3.3$, when the density field is overall smoother and much of the gas settles into a stable, hot atmosphere. 
This hot component forms through accretion shocks as well as through repeated cycles of AGN heating.
The cold filaments that connect the cosmic web to the central galaxy at $z \, = \, 6.2$ are more clearly disrupted at $z \,=\, 3.3$. We find that such filaments become disrupted between $z \,=\, 5.3$ and $z \,=\, 3.9$ for all haloes including the halo without AGN feedback. 

Due to its compact size, the quasar host galaxy is barely noticeable in the density projection at $z \, = \, 6.2$. In contrast, at $z \, = \, 3.3$, it clearly takes the shape of a larger disc, seen edge-on in the inset. The galaxy be seen as a cold, compact, disc surrounded by a large mass of cold, dense gas. 

Just through visual inspection, it is clear that the targeted galaxy experiences substantial size growth, as illustrated by the stellar surface density maps in the bottom panels of Fig.~\ref{fig:panels}. 
In the top row, we show the stellar distribution in our default setup (with AGN feedback) at $z \,=\, 6.2$ on the left-hand side, and at $z \,=\, 3.3$ on the right-hand side.
While the galaxy appears as a flattened spheroid of $\sim \rm kpc$ radius at $z \,=\, 6.2$, it has a size $\gtrsim 5 \, \rm kpc$ at $z \, = \, 3.3$.

The bottom row shows the results for the simulation without AGN feedback.
In contrast to our default simulation, the galaxy appears to experience less significant size growth and its surface density remains high and centrally concentrated. The maximum stellar surface density for the galaxy without AGN feedback is $2.3 \times 10^{6} \, \rm M_{\odot} pc^{-2}$ which is $\approx 66$ times higher than the maximum found for the galaxy with AGN feedback. 

The contours in the projected stellar maps show the surface density of the stars that make up the `compact bulge' at $z \, = \, 6.2$. Each stellar particle has a unique ID that allows it to be traced back- and forward in time in the simulation. We select the stars inside the stellar half mass radius (1.2 kpc) at $z = 6.2$ as proxy for `compact bulge' stars, extract their IDs and identify their location at $z = 3.3$. The contours correspond to $5, 100, 700 \, \rm M_{\odot} \, pc^{-2}$ surface density levels, where the thickest contour corresponds to the highest surface density. At $z \,=\, 6.2$, the `compact bulge' stars are, by construction, concentrated within the stellar half mass radius of $\approx 1$ kpc and there is little difference between simulations with and without AGN feedback.
While, at $z = 3.3$ the bulge stars in the galaxy without AGN feedback are less concentrated than at $z \,=\, 6.2$, they remain within a radius of $\approx 3$ kpc, a factor three larger than at $z \, = \, 6.2$. 
More strikingly, in the galaxy with AGN feedback, the `compact bulge' stars are significantly more spatially extended. The radius of the outer contour is $\approx 8$ kpc. 

Thus, despite the remarkable compactness of quasar host galaxies at $z \,=\, 6.2$, some process leads to the expansion of the `compact bulge' and the migration of its stellar populations to the outskirts of the galaxy. We have shown that this process is much more significant in the presence of AGN feedback, which, as explained in Section \ref{sec:numerical setup}, operates solely via thermal, `quasar mode' feedback at all regimes. We thus now explore how quasar-driven outflows shape the evolution of the quasar host galaxies. 

\subsection{The impact of AGN feedback on the properties of the quasar host galaxy}
\label{sec:AGN feedback}

\begin{figure}
\includegraphics[width=\columnwidth]{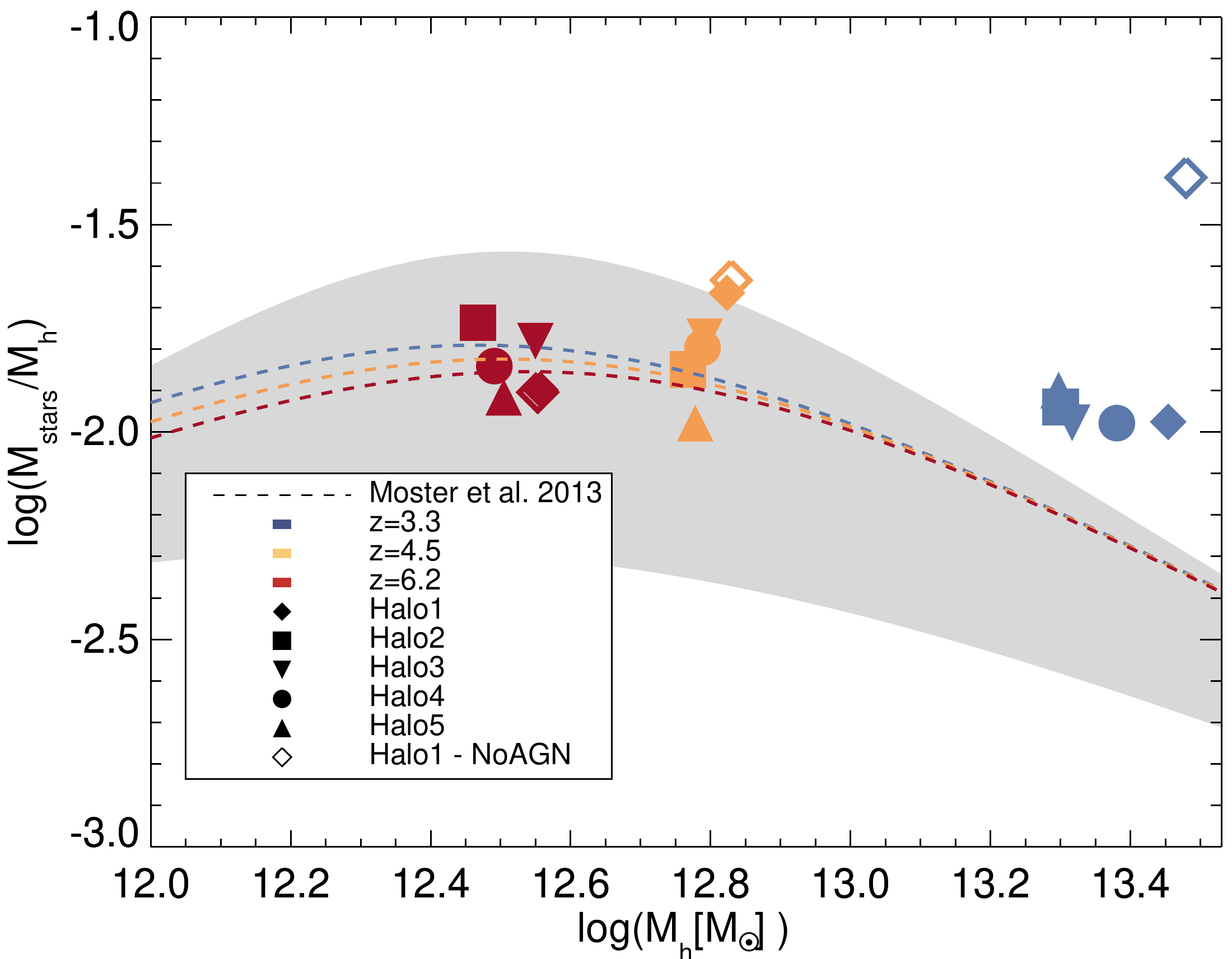}
    \caption{Total stellar mass-halo mass ratio versus halo mass for the five haloes shown with the abundance matching constraints of \citet{Moster_2013} for $z \,=\, 6.2$, $z \,=\, 4.5$ and $z \,=\, 3.3$ (shown with dashed lines). The colour-coded symbols show the redshift evolution of the haloes. The shaded area shows the plausibility range of the stellar mass-halo mass at $z \,=\, 3.3$ as in \citet{Moster_2013}. 
    There is good agreement between the simulated galaxies and the abundance matching constraints, though the simulated galaxies lie somewhat above the \citet{Moster_2013} relation at $z \, = \, 3.3$. If AGN feedback is excluded (open symbols), then the discrepancy between simulations and abundance matching constraints widens significantly.
    Note that at $z \,=\, 6.2$ the symbols of Halo1 and Halo1-NoAGN overlap.}
    \label{fig:abundance matching}
\end{figure}

\begin{figure*}
\includegraphics[width=\textwidth]{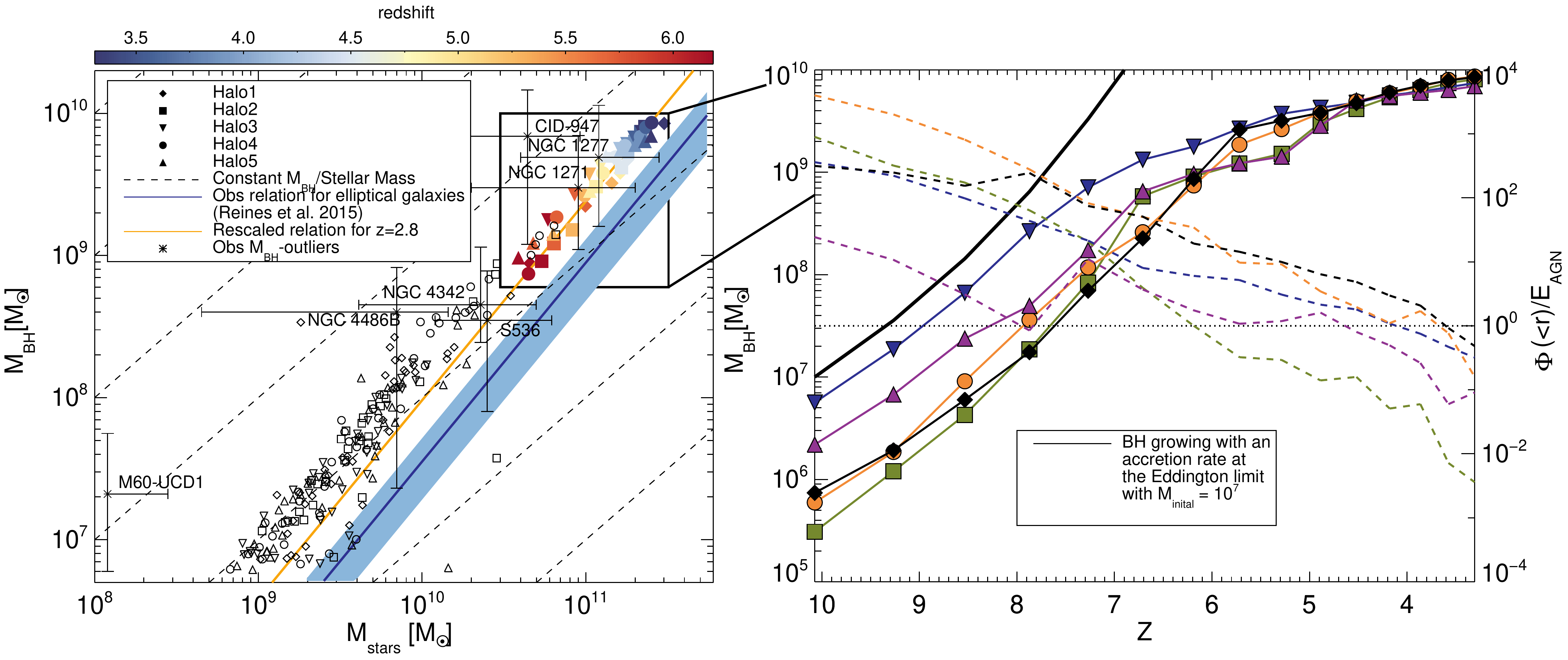}
    \caption{The left-hand panel shows the black hole mass versus total stellar mass for all haloes compared to the relation for elliptical galaxies of \citet{Reines_2015}. The blue-shaded region indicates the intrinsic scatter around the observed relation. We also re-scale this relation to higher redshift with the $(1 + z)^{0.68}$ scaling suggested by \citet{Merloni_2010}. The orange line indicates the re-scaled relation at $z \,=\, 2.8$. Various observed $M_{\rm BH}-M_{\rm stars}$ outlier galaxies are indicated with black stars (see text for details and references). The black open symbols refer to other black holes selected from every simulation at every snapshot. They give the values of 7 other massive black holes in every simulated environment taken at the same redshifts as the colour-coded ones. The right-hand panel shows the evolution mass of the most massive black holes in each simulation (left axis). The black line shows the black hole mass expected from constant Eddington-limited accretion, starting from an initial mass of $10^{7} \rm M_{\odot}$ at $z \, = \, 10$. The dashed lines show the ratio between the total gravitational potential, measured within 1 kpc, and the total feedback energy released (right axis).}
    \label{fig:Ms_Mbh}
\end{figure*}

Here, we investigate how strongly AGN feedback influences the evolution of quasar host galaxies and their host haloes. 

\subsubsection{Stellar mass vs. halo mass}

We link the stellar mass in the halo to the halo mass and compare these with existing abundance matching constraints. In Fig. \ref{fig:abundance matching} we plot the stellar mass-halo mass ratio against halo mass. The colour-coded symbols show the evolution of the haloes with redshift and the coloured lines indicate the redshift dependent relation found by \cite{Moster_2013} for $z \,=\, 6.2$, $z \,=\, 4.5$ and $z \,=\, 3.3$. For the stellar mass and total halo mass, we use the values as predicted by {\sc SUBFIND}.

We see that the simulations match the observational constraints well at $z \,=\, 6.2$ and $z \,=\, 4.5$. This agreement, however, breaks down at $z \,=\, 3.3$ (blue coloured symbols) and it is clear that our massive galaxies form stars too efficiently, when compared to the abundance matching constraints. However, the stellar mass fractions of the simulated haloes are in much better agreement with the observed relation than in the simulation without AGN feedback.
In the latter, the halo forms an order of magnitude more stars than expected based on the abundance matching constraints, while we see a disparity of only a factor two when we include AGN feedback. 

Thus, up until the later stages of our simulations, galaxy stellar masses appear reasonable and, if anything, AGN feedback appears to be too weak.
This is not surprising as continuous thermal energy injection is known to suffer from numerical radiative losses at the resolution typically achieved by even `zoom-in' cosmological simulations \citep{Booth_2009, Weinberger_2018}. The main point here, however, is that if AGN feedback is indeed responsible for the expansion of the galaxies' `compact bulges', this is unlikely to be because AGN feedback is unrealistically strong. 

\subsubsection{Black hole mass vs. stellar mass}

Another relation that tests our implemented feedback model is the black hole mass \-- stellar mass relation. Fig. \ref{fig:Ms_Mbh} shows the black hole mass as a function of total stellar mass in our various simulated haloes. The colour-coded symbols show the redshift evolution of the most massive black holes located in the nuclei of the most massive galaxies in each simulation. The black symbols give the values of 7 other massive black holes in every simulated environment and these are taken at the same redshift as the colour-coded ones. 

While the slope is in good agreement with the local relation, the normalisation appears to be higher by a factor of $\approx 3$. It has been claimed that observed high redshift black holes are, typically, over-massive at fixed galaxy mass (e.g., \citealt{Volonteri_2011}) and that there can be significant redshift evolution in various $\rm M_{\rm BH}-$ galaxy relationships. \citet{Merloni_2010} propose that $M_{\rm BH} - \rm M_{\rm stars}$ evolves with redshift as $\propto (1 + z)^{0.68}$, while \citet{Decarli_2010} suggest a different scaling $\propto (1 + z)^{0.28}$. If we rescale the relation found by \citet{Reines_2015} using the scaling found by \citet{Merloni_2010} (shown in orange in Fig. \ref{fig:Ms_Mbh}), we obtain better agreement between the simulations and the observations.
We note that the normalisation of the $M_{\rm BH} - \rm M_{\star}$ relation evolves in various other state-of-the-art simulations \citep{Barber_2016, Sijacki_2015}. 

$M_{\rm BH}$ and $M_{\rm stars}$ estimates for various observed galaxies with over-massive BHs are also shown in Fig. \ref{fig:Ms_Mbh} for reference. Values were taken from \citet{Seth_2014}, \citet{Saglia_2016}, \citet{McConnell_2013}, \citet{vanLoon_2015}, \citet{Trakhtenbrot_2015}, \citet{Walsh_2015} and \citet{Walsh_2016} for M60-UCD1, NGC 4486B, NGC 4342, S536, CID-947, NGC 1271, and NGC 1277, respectively\footnote{Note that for NGC 4486B, NGC 4342 and NGC 1271, we plot stellar bulge mass since total stellar mass is not available.}. We see that the simulated black holes compare well with the observed over-massive black holes and suggest that the simulated environments could thus be the birth place of such nearby observed over-massive black holes.

\subsubsection{Black hole growth}

Here, we investigate how the central black holes evolve before and after their bright quasar phase at $z \, = \, 6$, In the right-hand panel of Fig. \ref{fig:Ms_Mbh}, we show the central black hole mass as a function of redshift, where the colour-coded symbols indicate the evolution for the different haloes. 

Each black hole starts from an initial seed of $10^5 h^{-1} \rm M_{\odot}$ placed in the massive halo at around $z \,=\, 15$. By $z \,=\, 10.1$, the central black holes have grown to masses $3.1 \times 10^5 \-- 5.6 \times 10^6$ $\rm M_{\odot}$ through a combination of mergers and accretion \citep[see][]{Sijacki_2009}. In the redshift range $z \,=\, 10 - 6$, we see exponential growth as the accretion rate is close to the Eddington limit almost all the time. This can be seen from the comparison with the black line which shows the expected mass for a black hole with a mass of $10^{7} \rm M_{\odot}$ growing at the Eddington rate starting at $z \, = \, 10$. It is clear that the slopes of the black hole masses in all our simulations are similar to that expected for Eddington-limited accretion at $z \gtrsim 7$.

Black hole growth slows down significantly below $z \,=\, 6$. This indicates that AGN feedback has become efficient. This can also be seen from the dashed lines in the right-hand panel of Fig. \ref{fig:Ms_Mbh} which show ratio between the evolution of the total gravitational potential of gas within 1 kpc divided and the cumulative amount of feedback energy released with redshift. We can see that below $z \,=\, 6$ the energy released by the AGN becomes comparable to the total gravitational potential in the inner kpc of the galaxy. AGN feedback has become strong enough to remove gas from the inner kpc. The feedback prevents the gas from accreting onto the black hole, thus slowing down the black hole growth. 

AGN feedback is a circular process; after gas is removed from the central region, new gas will cool and move into the galactic nucleus and eventually accrete onto the black hole, starting a new feedback cycle (\citealt{Costa_2014}). These short bursts of accretion allow the black hole to continue growing, though at modest rates. By $z \, = \, 3.3$, the central black holes in our simulations reach masses in the range $6.9 \-- 8.6 \, \times 10^9 \rm M_{\odot}$. 

\subsubsection{Morphology}
\label{sec:Morphology}

In order to quantify the morphology of the compact stellar component that forms at the centre of the quasar host galaxies, we evaluate the fraction of kinetic energy invested in motion which is co-rotating with respect to the halo (\citealt{Correa_2017}). We evaluate the $\kappa_{\rm rot}$ parameter, defined as

\begin{equation}
\kappa_{\rm rot} = \frac{K_{\rm rot}}{K} = \frac{1}{K}\sum_{i}^{r<30 \, \rm kpc} \frac{1}{2}m_i[L_{z,i}/m_iR_i]^2 \, ,
\end{equation}

where the sum is performed over all stellar particles within a spherical radius of $30 \, \rm kpc$ centred at the position of the most massive halo, $m_i$ is the stellar particle mass, $K =\sum_{i}^{r<30 \, \rm kpc} \frac{1}{2}m_iv_i^2 $ the total kinetic energy, $L_{z,i}$ the particle angular momentum along the direction of the total angular momentum of the stellar component inside the radius of 30 kpc and $R_i$ the distance to the centre of the galaxy. Following \citet{Correa_2017} we only take corotating stellar particles ($L_{z,i} > 0$) into account which is found to be a better measure of ordered rotation.
\citet{Correa_2017} find that $\kappa_{\rm co}$ is a reasonable measure of visual morphology and established that $\kappa_{\rm co} = 0.4$ can be used to separate disc-like galaxies ($\kappa_{\rm co} \geq 0.4$) from elliptical galaxies ($\kappa_{\rm co} < 0.4$). We thus use also $\kappa_{\rm co} = 0.4$ to separate the two populations. 

\begin{figure}
\includegraphics[width=\columnwidth]{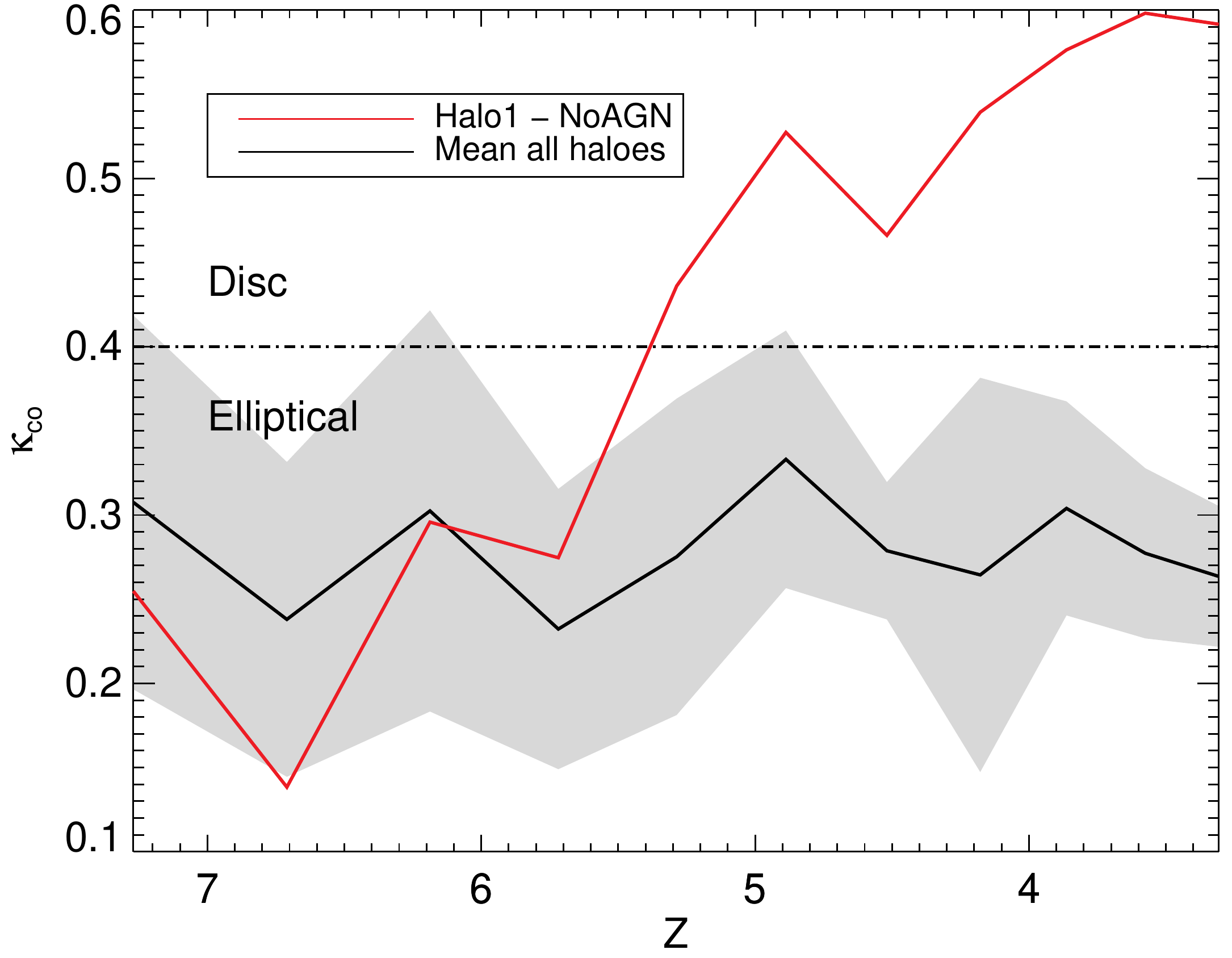}
    \caption{The evolution of the quasar host galaxy morphology as quantified by the $\kappa_{\rm co}$ parameter (see text). The solid black line shows the mean of all galaxies with AGN feedback, whereas the grey shaded region shows the standard deviation. The red solid line shows the evolution of the galaxy without AGN feedback. All quasar host galaxies are consistent with an elliptical morphology throughout the simulation. In the absence of AGN feedback, however, the massive galaxy develops a prominent stellar disc. From the standard deviation, we see that disc-like stellar structures form in some quasar host galaxies from time to time. These galaxies are, however, unable to maintain this structure.}
    \label{fig:kappa}
\end{figure}

Fig. \ref{fig:kappa} shows the redshift evolution of $\kappa_{\rm co}$ where the black line shows the mean evolution for all quasar host galaxies and the solid red line shows the evolution for the galaxy without AGN feedback. We confirm that all investigated galaxies start as dispersion-dominated ellipticals \citep[as in][]{Dubois_2012}. At $z \,=\, 3.3$, the galaxies with AGN feedback have a mean value of $\kappa_{\rm co} = 0.26$ and are thus in agreement with an elliptical morphology. 

In addition, we find that the evolution of the mean value for all galaxies indicates that the quasar host galaxies never develop pronounced discs. From the standard deviation shown as the grey shaded region in Fig. \ref{fig:kappa}, we see that disc-like stellar structures form in some quasar host galaxies. Specifically, the galaxy in Halo3 becomes disc-like at $z \,=\, 7.2$ and $z \,=\, 6.2$. Form $z \,=\, 6.2$, this galaxy stays elliptical. The galaxy in Halo1 also becomes disc-like at $z \,=\, 4.8$ but remains elliptical from $z \,=\, 4.5$. Quasar host galaxies are thus able to form disc-like stellar structures but do not maintain the disc-like structure. This in contrast to the galaxy without AGN feedback, which becomes disc-like at $z \,=\, 5.2$ and is able to maintain this structure reaching $\kappa_{\rm co} = 0.59$ at $z \,=\, 3.3$. The disc-like structure can also be seen from the stellar surface density maps, shown in the bottom panels of Fig. \ref{fig:panels}. The galaxy without AGN feedback gains this disc-like configuration at $z \,=\, 5.3$ and remains disc-like from then onwards.
A similar behaviour is seen in the HORIZON-AGN simulations in the case in which AGN feedback is excluded; late-time accretion of gas results in the formation of a disc while, if AGN feedback is active, the formation of the disc is prevented (\citealt{Dubois_2016}).

AGN feedback thus appears to ensure that the galaxies hosting quasars at $z \gtrsim 6$ contain spheroidal, dispersion-dominated stellar components.
We emphasize that the gas component, which evolves on Myr timescales, does have a disc-like morphology, but we find that this changes orientation rapidly as the central galaxy is bombarded by satellite galaxies as well as by a smoother inflow component.
The resulting angular momentum cancellation ultimately promotes the formation of a compact, dispersion-dominated bulge \citep{Dubois_2012}.

\begin{figure*}
\includegraphics[width=\textwidth]{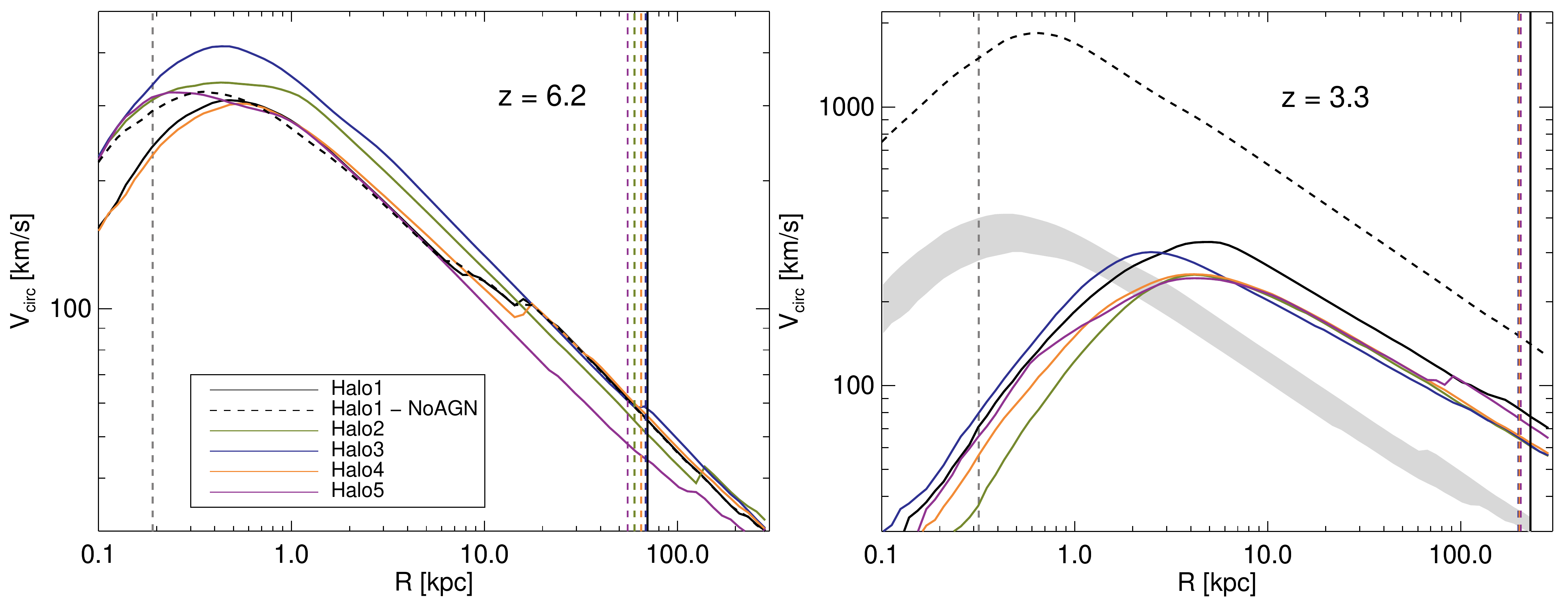}
\caption{Circular velocity profiles of the stellar component in our simulation sample at $z \,=\, 6.2$ in the left-hand panel and at $z \,=\, 3.3$ in the right-hand panel. The simulated haloes with AGN feedback are shown with the coloured solid lines and the simulated halo without AGN feedback is shown with the dashed black line. The vertical grey dashed line marks the location of the gravitational softening length. The vertical coloured dashed lines mark the location of the virial radii of the quasar hosting haloes. The vertical black solid line marks the location of the virial radius of the simulated halo without AGN feedback. The grey shaded region in the right-hand panel shows the range of circular velocities at $z \,=\, 6.2$ from the left-hand panel. While all quasar host galaxies are remarkably compact and tightly bound, with stellar circular velocities peaking at $300 \-- 500 \, \rm km \, s^{-1}$ at radii $\approx 500 \, \rm pc$ at $z \, = \, 6.2$, they become larger and less tightly bound at $z \, = \, 3.3$ (as long as AGN feedback operates).}
\label{fig:v_circ_at_z6}
\end{figure*}

\subsubsection{Circular velocity}
\label{sec:Circular velocity}

In Fig. \ref{fig:v_circ_at_z6} we show stellar circular velocities profiles for all our simulations at $z \,=\, 6.2$ and $z \,=\, 3.3$. The circular velocity is defined as
\begin{equation}
V_{\rm circ} = \sqrt{\frac{G M_{\rm stars}(<R)}{R}}\, ,
\label{eq:v_circ}
\end{equation}
where $M_{\rm stars}(<R)$ is the stellar mass enclosed within a radius $R$. 

\begin{figure}
\includegraphics[width=\columnwidth]{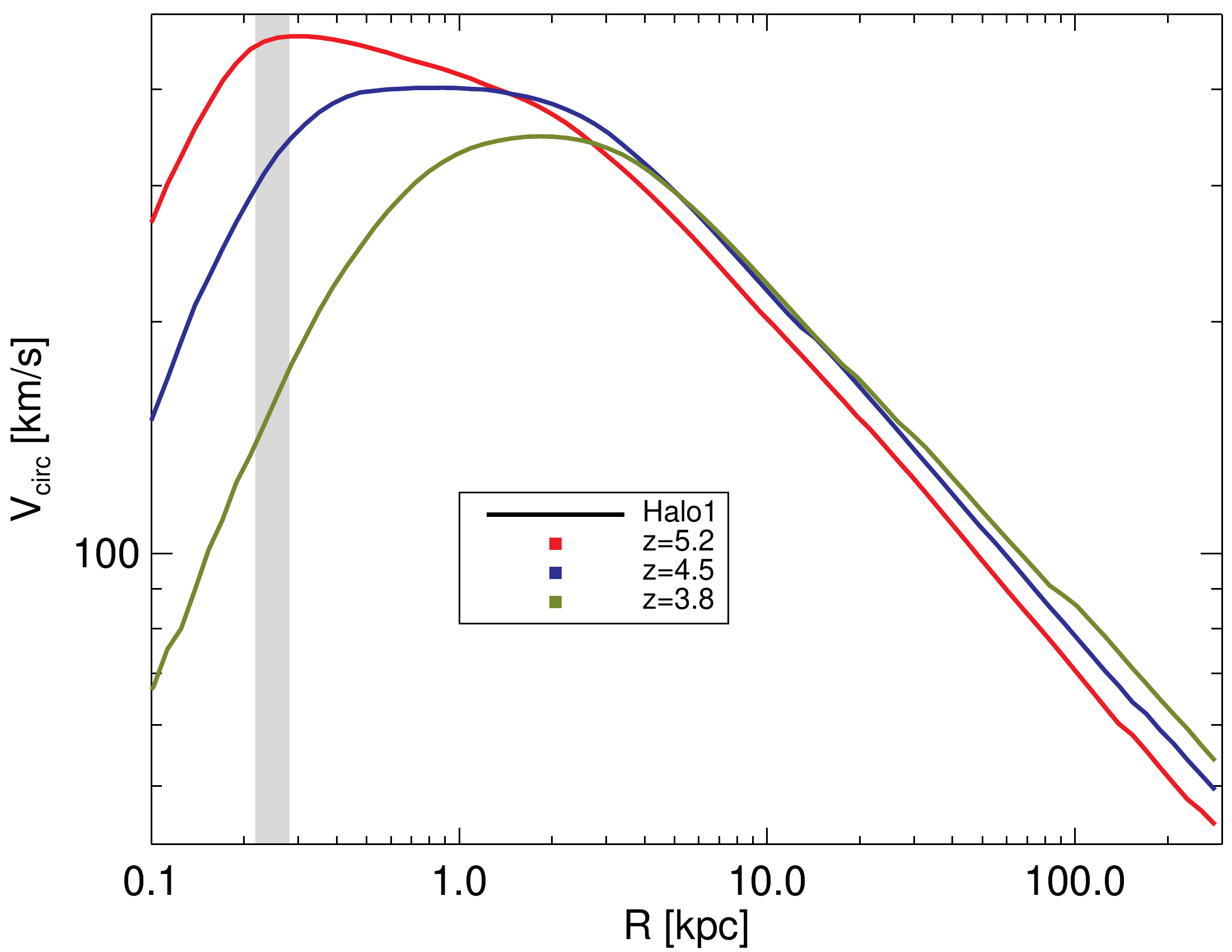}
\caption{The circular velocity profile of the stellar component for Halo1. The profile is shown at different redshifts with the coloured solid lines. The vertical grey area marks the location of the gravitational softening length for $z  \, = \, 5.2$ to $z  \, = \, 3.8$. As redshift decreases, the peak circular velocity drops and its location shifts to larger radii, indicating a gradual expansion of the quasar host galaxy.}
\label{fig:v_circ_z}
\end{figure}

At $z \, =\, 6.2$, all quasar host galaxies are, as we have seen in Fig.~\ref{fig:panels}, very compact.
If we consider the radial position of the maximum (stellar) circular velocity $R_{\rm max}$ as a proxy for bulge size, we find $R_{\rm max} \approx 0.2 \-- 0.5 \, \rm kpc$ with a mean of $\sim 0.4$ kpc which is roughly twice the size of the gravitational softening length. The compact bulges have peak circular velocities between 300 km s$^{-1}$ and 500 km s$^{-1}$. Stars make up between 40\% and 57\% of the total mass within the inner kpc where dark matter and gas make up between 10\% and 33\% of the total mass. The black hole only makes up $\sim 3$\% of the total mass.

In addition, we note that, at $z \,=\, 6.2$, the circular velocity profile of the simulation without AGN feedback is comparable to the circular velocity profiles found in the other five simulations. We find a peak circular velocity of $\sim 340$ km s$^{-1}$ for the halo without feedback, comparable to the mean of $\sim 350$ km s$^{-1}$ for haloes with AGN feedback. This, along with Figs~\ref{fig:panels} and~\ref{fig:abundance matching}, indicates that AGN feedback has not yet significantly impacted the structure of the galaxy by $z \,=\, 6.2$, despite the presence of vigorous quasar-driven outflows \citep{Costa_2015, Curtis_2016}.

At $z \,=\, 3.3$, all quasar host galaxies have, as we have seen for one of our simulations in Fig.~\ref{fig:panels}, become less compact. The radial position of the maximum circular velocity grows to $R_{\rm max} \approx 2.5 \-- 4.0 \, \rm kpc$, an order of magnitude higher than the gravitational softening length and $R_{\rm max}$ at $z \, =\, 6.2$. \emph{Thus, while a compact bulge is initially found in every of our quasar host galaxies, it expands significantly in all our simulations.}
The evolution of the stellar circular velocity for a single one of our systems can be seen in Fig. \ref{fig:v_circ_z}. The peak of the circular velocity profile shifts to larger radii and decreases as the redshift decreases. 

Importantly, the above statements only hold in simulations with AGN feedback.
In the simulation without AGN feedback, the compact bulge seen at $z \,=\, 6.2$ becomes more compact and more massive by $z \,=\, 3.3$; the maximum circular velocity increases from $300$ km s$^{-1}$ to $2000$ km s$^{-1}$. The peak value of the circular velocity stays at roughly the same radius of $\sim 0.3$ kpc. From Fig.~\ref{fig:panels} we do, however, observe some stellar migration which is likely due to numerical effects. This effect is further explored in Section \ref{sec:Mechanisms for size growth}.

\begin{figure*}
\includegraphics[width=\textwidth]{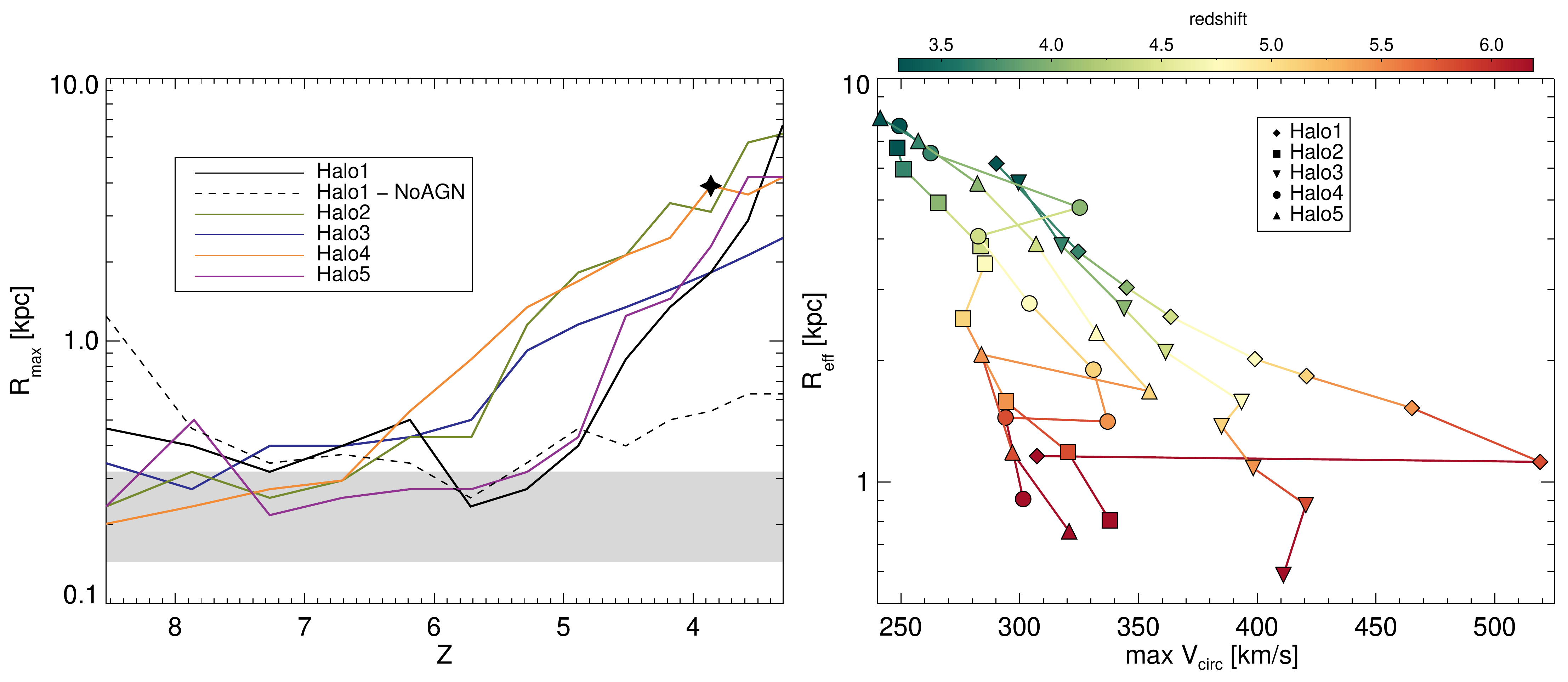}
    \caption{The left-hand panel shows the location of the maximum stellar circular velocity as a function of redshift for the six investigated galaxies. The solid lines show the evolution in our simulations with AGN feedback, while the dashed line shows the results for the simulation without black holes and AGN feedback. The grey band indicates the range of stellar gravitational softening lengths, which, since they are fixed in comoving coordinates, change with redshift in physical coordinate. At $z \, \approx \, 4$, the massive galaxy in Halo4 undergoes a major merger, which results in a double-peaked circular velocity profile; we have here taken the radius associated with the second peak. The right-hand panel shows the redshift evolution of the stellar half mass radius as a function of the maximum stellar circular velocity. We see that for $z \, \lesssim \, 6$, all quasar host galaxies experience size growth, while their maximum circular velocity drops.}
    \label{fig:size_growth}
\end{figure*}

In conclusion, we thus confirm that our implemented AGN feedback model is reasonable and does a reasonable job in preventing the overproduction of stars in massive galaxies (but is still likely too weak, especially at later times). The simulated black holes also compare well with the observed over-massive black holes. When examining the morphology of the quasar host galaxies, we find that these stay elliptical throughout the simulation. The galaxy without AGN feedback, on the other hand, gets a disc-like configuration. We also find that, while all quasar host galaxies host highly compact stellar bulges at $z \, = \, 6$, they all experience significant expansion with time. The galaxy without AGN feedback maintains its compact bulge. In the following section, we investigate whether AGN feedback is directly responsible for the expansion of the bulge. 

\subsection{Growing galaxies with AGN-driven outflows}
\label{sec:Evolution Bulge halo1}

Size is not a well-defined parameter for massive galaxies with extended stellar mass distributions, which makes it hard to compare our size measurements to other simulations and observations. The effective radius used in observations depends, for example, on resolution, filter and the adopted model for the light profile. We have therefore adopted the position of the peak of the stellar component circular velocity ($R_{\rm max}$) as a simple proxy for radius. This radius encloses approximately $15 \-- 30$ \% of the stellar mass within the virial radius. We also use the stellar half-mass radius as defined by {\sc SUBFIND} $R_{\rm eff}$ as proxy for radius which, obviously, encloses $50$ \% of the stellar mass within the subhalo. 

\subsubsection{Size growth across time}
\label{sec:Evolution circular velocity}

The redshift evolution of the size of the targeted galaxies, here defined as the radius of the stellar circular velocity peak is shown on the left-hand panel of Fig. \ref{fig:size_growth} with solid lines for simulations including black holes and AGN feedback and a dashed line for the simulation without black holes and AGN feedback. The grey band in the figure indicates the range over which the gravitational softening lengths change over time in our simulations.

The targeted galaxies start with roughly the same size, as we have seen in the previous section, ranging from $\approx 0.2$ kpc to $\approx 0.4$ kpc at $z \,\approx\, 7$. In simulations which include AGN feedback, the massive galaxies start growing in size after $z \,=\, 7$. The sizes of the galaxies in Halo2 and Halo4 start to grow rapidly after $z \,=\, 7$ whereas the other galaxies initially grow only slowly. The galaxy in Halo3 starts to grow faster after $z \,=\, 5.7$ and the galaxy in Halo5 starts to grow rapidly in size after $z \,=\, 4.8$. The galaxy in Halo1-NoAGN grows little, from $\approx 0.5$ kpc to $\approx 0.6$ kpc; it has roughly 10 \% of the mean size of the other galaxies at $z \,=\, 3.3$ which is $\approx 5 \, \rm kpc$.

In the right-hand panel of Fig. \ref{fig:size_growth}, we plot the stellar half mass radius as a function of the maximum value of the circular velocity for all simulations with AGN feedback. We see that, as the quasar host galaxies grow in size, their maximum stellar circular velocity drops, indicating that the galaxies gradually become less tightly bound.
The peak circular velocity and the stellar half mass radius have similar values for all haloes at $z \,=\, 6.2$ and all galaxies end up having roughly the same half mass radius of $6 \-- 9$ $\rm  kpc$ at $z \,=\, 3.3$. 
While galaxy growth is particularly rapid at $z \gtrsim 4$, it slows down at lower redshift, when the maximum circular velocity reaches values of$\approx 250 \, \rm  km \, s^{-1}$. Halo5 is simulated down to $z \,=\, 2.6$ and continues to show size growth at a rate similar to that seen in all targeted haloes at $z < 4$.
One striking feature in the right-hand panel of Fig. \ref{fig:size_growth} is the sudden increase of the peak circular velocity in Halo1 between $z \,=\, 6.2$ and $z \,=\, 5.7$. Stellar images and circular velocity profiles show that a merger is happening at this point which increases the circular velocity. Afterwards, the galaxy shows the same evolution as the other quasar hosting galaxies, the galaxy grows in size and its maximum stellar circular velocity drops gradually.
We also note that for Halo1-NoAGN, the maximum circular velocity increases rapidly to extreme values $\lesssim 2000 \, \rm km \, s^{-1}$, while the galaxy size remains close to $\approx 1 \, \rm kpc$ (see Fig.~\ref{fig:v_circ_at_z6}). 

While it is clear that the `puffing-up' of the stellar component appears to be connected to the presence of an AGN, as found in various previous studies \citep{Peirani_2017, Costa_2018, Choi_2018}, we have not yet shown that AGN feedback plays a direct role on the expansion of the compact bulges. For instance, it could act simply by reducing the gas fraction of the progenitor galaxies and thereby increasing the number of dry mergers undergone by the progenitors. As mentioned in Section \ref{sec:Introduction}, this is the prevailing explanation for the size growth of high redshift massive compact ellipticals. In the following sections, we show that, in our simulations, AGN feedback plays quite a direct role in galaxy size growth.

In order to understand how AGN feedback may provoke galaxy growth, we have performed a new simulation identical to Halo1, but using a larger number of simulation outputs with a $z \approx 0.02$ spacing, which is comparable to the dynamical time-scale measured at $\approx 1$ kpc. The simulation is run from $z \,=\, 6.7$ to $z \,=\, 4.5$, the time range where we saw the most intense size growth.
In addition, we also perform another simulation like Halo1, but for which we `switch-off' black hole accretion and AGN feedback at $z \, = \, 5.2$.
We analyse the results of these simulations in the next section.

\subsubsection{Fluctuations in the gravitational potential}
\label{sec:Fluctuating potential}
\begin{figure*}
    \centering
    \subfigure[]
    {
        \includegraphics[width=1.0\columnwidth]{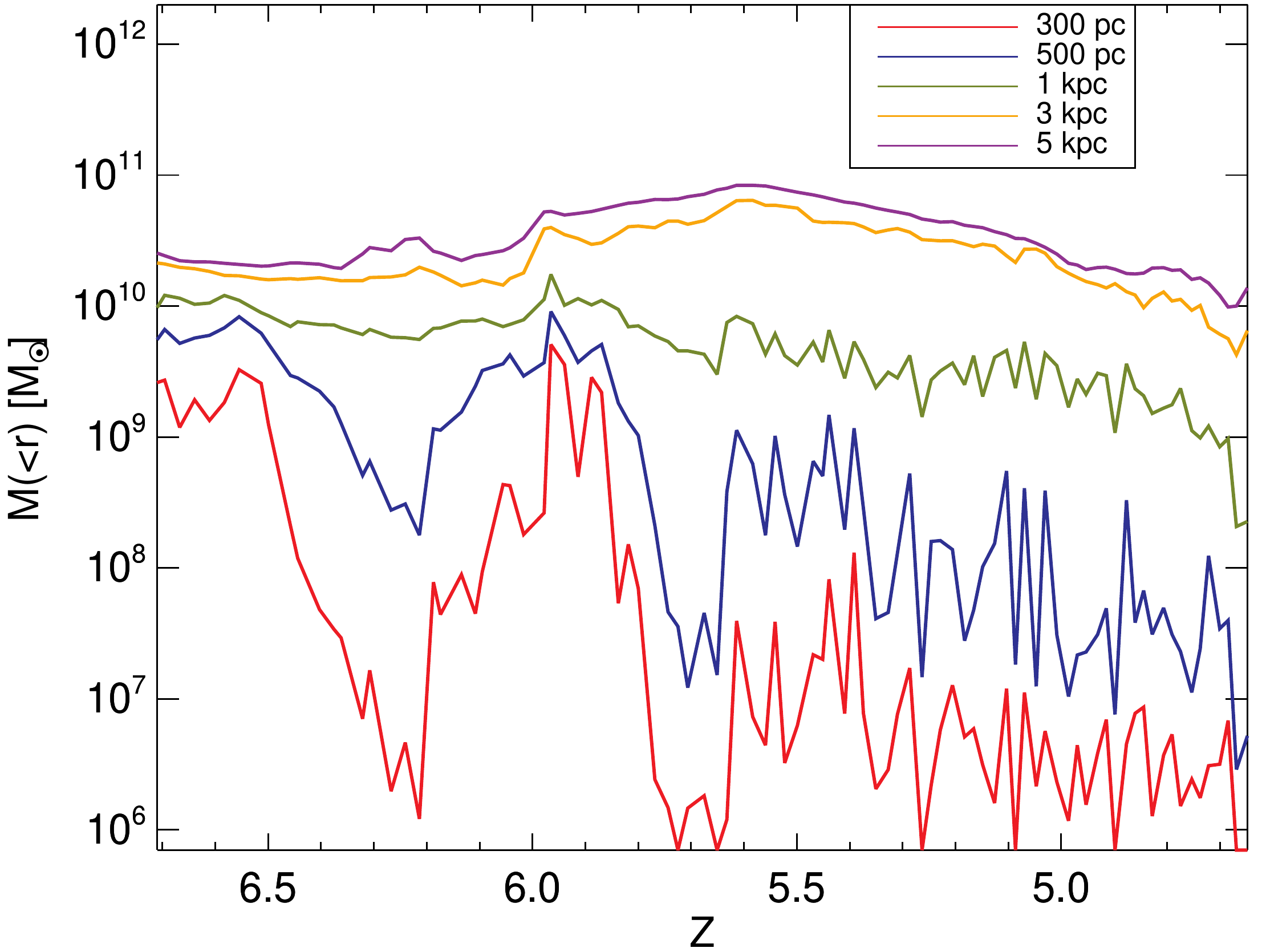}
        \label{fig:enclosed_m_1}
    }
    \subfigure[]
    {
        \includegraphics[width=1.0\columnwidth]{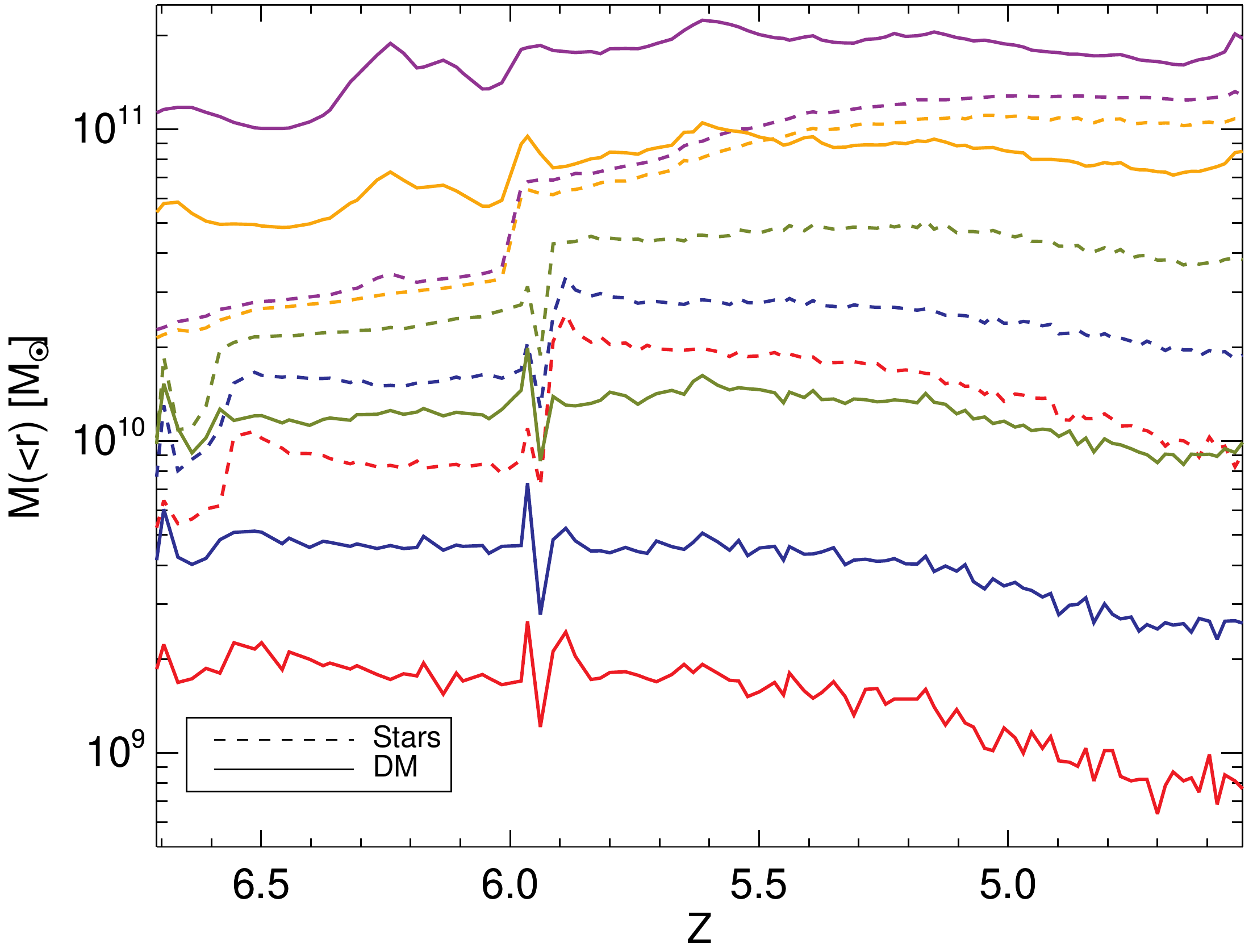}
        \label{fig:enclosed_m_2}
    }
    \subfigure[]
    {
        \includegraphics[width=1.0\columnwidth]{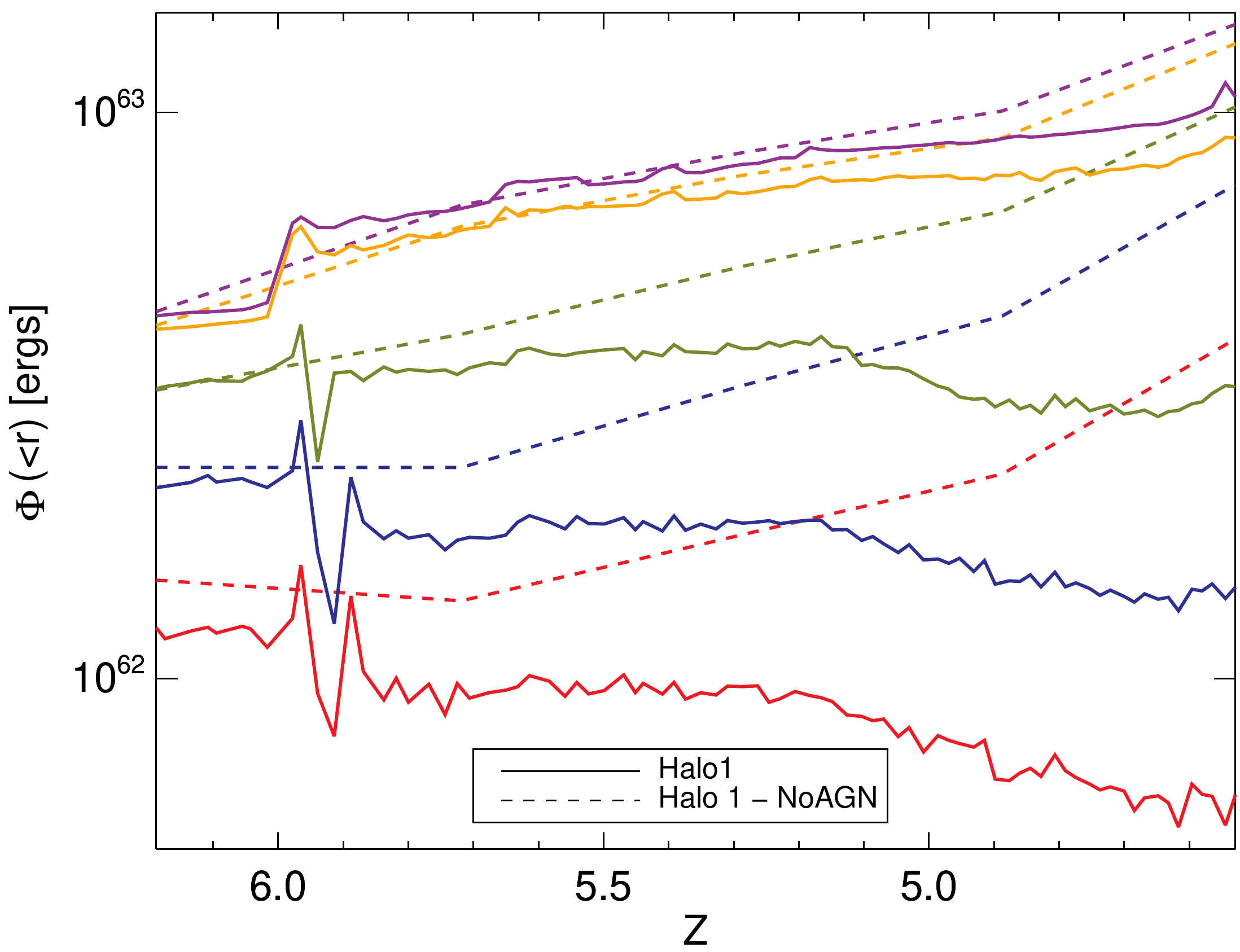}
        \label{fig:grav_pot_bulge_stars}
    }
    \caption{Panels \subref{fig:enclosed_m_1} and \subref{fig:enclosed_m_2} show the enclosed mass for different radii where \subref{fig:enclosed_m_1} shows the gas component and \subref{fig:enclosed_m_2} the stellar and dark matter component indicated with, respectively, solid and dashed lines. Clearly, the enclosed gas mass becomes highly variable after $z \, \approx \, 5.7$ within $\sim \rm kpc$ scales, approximately when the released AGN feedback energy becomes comparable to the galaxy binding energy. From this time onward, the central stellar and dark matter mass also starts to drop. The redshift evolution of the total gravitational potential for bulge stars (selected at $z \, = \, 6.2$) within different radii is shown in \subref{fig:grav_pot_bulge_stars}. In the absence of AGN feedback (dashed lines), the gravitational potential of `bulge stars' increases, in contrast to all other simulations. 
    }
\end{figure*}

Fig. \ref{fig:enclosed_m_1} shows the enclosed gas mass in Halo1 at different radial scales, as indicated by different colours. AGN activity becomes strong enough to regulate the central gas reservoir already at $z \,=\, 6.5$, when we see a drop in the enclosed mass within a radius of $500 \, \rm pc$. At $z \lesssim 6$, we see rapid, $\approx$ $1 \, \rm Myr$, and strong fluctuations in the enclosed mass. The pronounced peak seen at $z \,\approx\, 6$ is caused by a merger which temporarily supplies the galaxy with fresh gas. The gas mass inside $1 \, \rm kpc$ then starts to decrease below $z \,=\, 5.7$ more systematically, while the fluctuations in the enclosed gas mass start becoming noticeable at even larger radii of up to $3 \, \rm kpc$. 

The enclosed dark matter and stellar masses also undergo small fluctuations in response to the gas motions as can be seen from Fig. \ref{fig:enclosed_m_2}. The enclosed collisionless component mass also show a significant and gradual decline below $z \, = \, 6$ which occurs mainly within scales $\lesssim 3 \, \rm kpc$ and is not seen at larger scales. This suggests the gradual expansion of stellar and dark matter within those scales, in good match to what was seen in the circular velocity profiles discussed in the previous section.

We also measure the actual gravitational potential of `compact bulge' stars, which we here select as those stellar particles located within a radius of 3 kpc at $z \,=\, 6.2$. We track the stellar particles using their IDs and calculate their total gravitational potential $\Phi \, = \, - \sum_{i} \phi_{i}$, where the sum is performed over the gravitational potential $\phi_{\rm i}$ of each bulge stellar particle, within different radii. As can be seen in Fig. \ref{fig:grav_pot_bulge_stars}, we find the expected fluctuations starting from $z \,=\, 5.5$ in the gravitational potential of the bulge stars within radii of 1 kpc, and from $z \,=\, 5.2$ onwards, we also see the fluctuations for radii < 3 kpc. As the collisionless component migrates outwards, the gravitational potential also drops with time.

In the absence of AGN feedback (dashed lines in Fig. \ref{fig:grav_pot_bulge_stars}), the gravitational potential of `bulge stars' increases, in contrast to the simulation with AGN feedback.

\subsubsection{Ex/in situ stars}
\label{sec:Ex/in situ stars}

\begin{figure*}
\includegraphics[width=\textwidth]{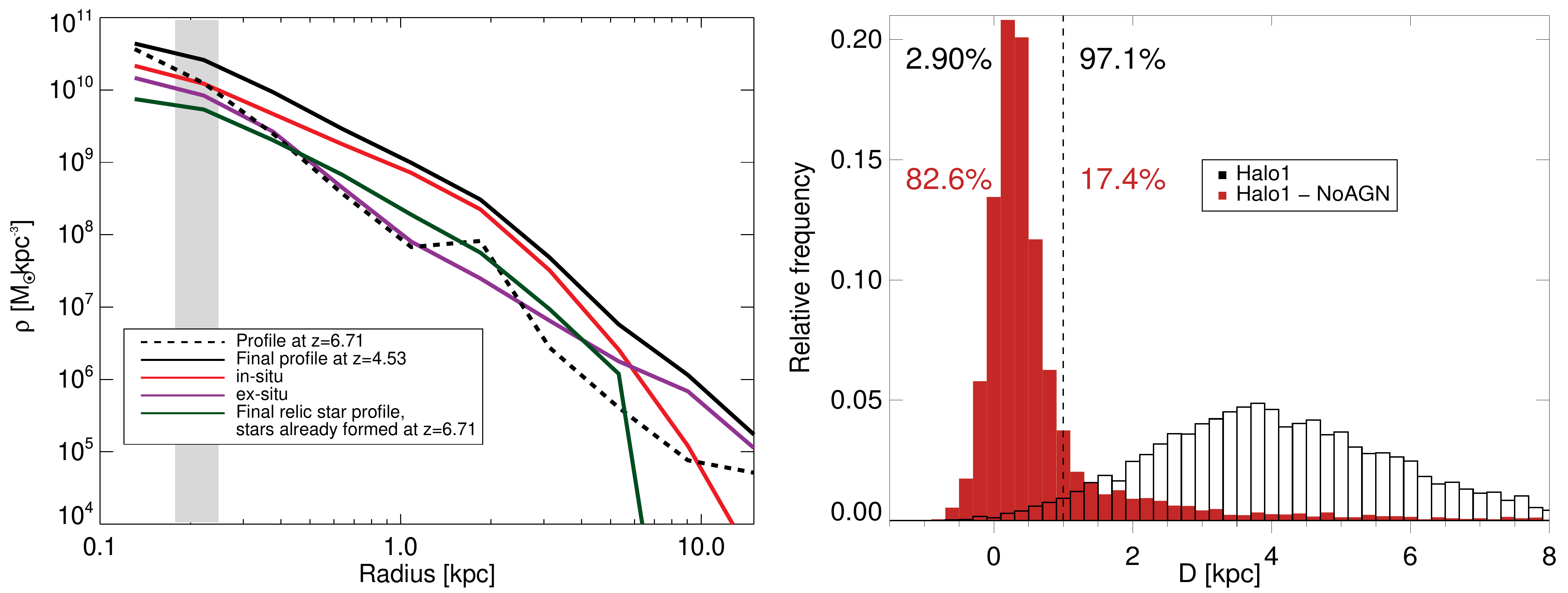}
    \caption{The left-hand panel shows the contribution to the stellar density profile of in-, ex-situ and of the bulge `relic' stars that existed at $z \, = \, 6.7$ already. Also shown is the total stellar density profile at $z \, = \, 6.7$ (dashed black line) and at $z \,=\, 4.5$ (solid black line). The right-hand panel shows the stellar radial distance travelled since $z \, = \, 6.7$ measured for star particles that reside within $r < 1 \, \rm kpc$ at $z \, = \, 6.7$ for the galaxy in Halo1 (black histogram) and the galaxy in the re-run simulation of Halo1 without AGN feedback (red histogram). The values give the percentage of star particles with $D$ greater or smaller than $1 \, \rm kpc$. When AGN feedback operates, most stars that make-up the initially compact quasar host galaxy migrate outwards by several kpc. If AGN feedback is neglected, outwardly stellar migration is far less significant.
    }
    \label{fig:in/ex}
\end{figure*}

The size growth we see in our simulations is, in part, driven by the expansion of the old stellar component which already formed before $z \,=\, 6.2$ (see bottom panels in Fig. \ref{fig:panels}). The accretion of stars that formed outside the galaxy (`ex-situ' stellar populations) could, however, also contribute to the overall size growth, as it builds up the outer layers of the quasar host galaxy.

To quantify the contribution of different populations to the size growth, we follow the stellar particles with their IDs and classify them based on their formation radius. We distinguish between `ex-situ', `in-situ' and `relic' stars. 
We define ex-situ stars as those which form outside $R_{\rm thresh} \, = \,  3 \times$ the stellar half mass radius, as defined by {\sc SUBFIND}. This radius is measured at every snapshot and changes thus with redshift. The in-situ stars are defined as those which form within $R_{\rm thresh}$. This radial threshold corresponds to $\approx$ 5\% of the virial radius which is somewhat smaller than the radius used in \citet{Oser_2010} and \citet{Frigo_2018}, which is 10\% of the virial radius. The choice of radial threshold, however, does not significantly affect our results as long as it lies in the range 5-15\% of the virial radius. 
Stars at $z \,=\, 4.5$ within $R_{\rm thresh}$ that already existed at $z \, = \, 6.7$ within $R_{\rm thresh}$ are defined as relic stars. 
We select an initial redshift of $z \, = \, 6.7$ here, because a highly tightly bound stellar bulge has already formed by that time. 

The density profile evaluated at $z \, = \, 4.5$ shown in Fig. \ref{fig:in/ex} with the solid black line consists of three different components: in-situ stars, ex-situ stars and relic stars. 
We find that the in-situ stellar population dominates the stellar mass at scales $\lesssim 10 \, \rm kpc$, making up $\approx 55$\% of the final mass, and, in particular, within the innermost few kpc, where size growth takes place.
In contrast, ex-situ stars make up $\approx 23$\% of the total stellar mass within $R_{\rm thresh}$, while the relic stars make up $\approx 18$\% of the final mass within $R_{\rm thresh}$. $\approx 4$\% of the stellar mass is not found using any these definitions, because they lie outside $R_{\rm thresh}$ at $z \,=\, 4.5$. The ex-situ stellar population is more concentrated than the in-situ stellar population. We find that early accreted ex-situ stars make up this concentrated central part. The contribution of ex-situ stars to the total final stellar mass is thus small compared to the contribution of in-situ stars, but comparable to the contribution of relic stars.
Note that if we select relic stars to be those which are in place at $z \, = \, 5.5$, instead of $z \, = \, 6.7$, we find that they contribute $\approx 60$\% to the total stellar mass within $R_{\rm thresh}$.

To see how different stellar populations contribute to the size growth, we check the difference between the formation radius and their radius at $z \approx 4.5$ ($\Delta R$ = formation radius - final radius). We see that $\approx 39$\% of the in-situ stars migrate inwards and have a greater formation radius than their final radius. In contrast, $\approx 61$\% of the in-situ formed stars migrate outwards where $\approx 1$\% of the outward migrated stars have a significantly bigger final radius ($\Delta R$ < -5 kpc). 

The relic stars also mostly move out to greater radii with time as can be seen in Fig. \ref{fig:in/ex} from the difference between the black dashed line, the profile at$z \, = \, 6.7$, and the green solid line, the profile at $z \,=\, 4.5$. About $86$\% of the relic stars move to larger radii and $\approx 5$\% of the outward-migrating stars have a significantly bigger final radius ($\Delta R$ < -5 kpc). 

In order to quantify how relic stars are affected by size growth, we use a similar approach as \citet{Choi_2018}. First, we select all stars within the stellar half mass radius at $z \, = \, 6.7$. We then trace these stars down to $z \,=\, 3.3$ and measure the radial distance $D$ they have traveled.
The red and black histograms in the right-hand panel of Fig. \ref{fig:in/ex} show the stellar radial migration distance from $z \, = \, 6.7$ to $z \,=\, 3.3$ for, respectively, the galaxy without and with AGN feedback in Halo1.

Most stars migrate radially outwards since $z \, = \, 6.7$ for the galaxy with AGN feedback, 97.1\% of the selected stars migrate more than 1 kpc outwards. The difference with the galaxy without AGN feedback is clear, only 17.4\% of the stars migrate radially outward more than 1 kpc. 82.6\% of star particles do not migrate further than 1 kpc since $z \, = \, 6.7$. The majority of star particles which constituted the core at $z \, = \, 6.7$ still are within the central region in the galaxy without AGN feedback. 

In this section we investigated three different stellar populations and showed how they contributed to the size growth observed in Section \ref{sec:Circular velocity}. We find that the contribution of in-situ stars to the density profile at $z \,=\, 4.5$ dominates. In-situ and relic stars are found to contribute most to the size growth as they migrate mostly outwards. We also show that relic stars migrate outwards in the galaxy in Halo1 which is not the case for the galaxy in the re-run simulation of Halo1 without AGN feedback. In addition, we observe that the density profile of the galaxy shown on the left-hand panel in Fig. \ref{fig:in/ex} experiences a flattening because of the outward-migrating stars. The black dash-dotted line (density profile at $z \, = \, 6.7$) shows an overall steep profile whereas the black solid line (density profile at $z \,=\, 4.5$) shows the start of the formation of a core in the profile. For the galaxy without AGN feedback we observe that the dark matter and the stellar density profiles only steepen with decreasing redshift. This will be further discussed in Section \ref{sec:cusp/core}. 

\section{Discussion}
\label{sec:Discussion}
\subsection{Growing massive galaxies through AGN feedback}
We have unambiguously shown that AGN feedback is directly responsible for the gradual expansion of the compact stellar bulges that form in massive galaxies at $z > 3$.
Size growth driven by AGN feedback in our simulations begins roughly when the energy injected by the AGN becomes comparable to the binding energy of the compact bulges.

While they have a scale radius of a few $100 \, \rm pc$ at $z \approx 6$, a time at which black hole growth is efficient and proceeds close to the Eddington rate, the compact bulges gain sizes an order of magnitude larger by $z \,=\, 3$ as long as AGN feedback operates.
The `puffing-up' of the stellar component has been previously described in the literature \citep{Fan_2008, RagoneFigueroa_2011, Lapi_2018} and has also been seen in cosmological simulations \citep{Peirani_2017, Peirani_2018, Costa_2018, Choi_2018} comparing the properties of galaxies with and without AGN feedback. We note that in those studies, fluctuations in the gravitational potential driven by outflows had been suggested as the main agent of galaxy growth, but this had not been explicitly shown. That AGN feedback drives sufficiently strong gravitational potential fluctuations has, however, been shown using idealised simulations \citep{Martizzi_2013}. The above described mechanism will be further explored in Section \ref{sec:Mechanisms for size growth}.

Based on our results, we find that AGN feedback may result in significant size growth in galaxies that have recently hosted intense quasar activity. We suggest that massive galaxy growth occurs in different phases: (i) At high-z, these systems initially go through a phase in which gas is accreted onto the nucleus at high rates, powering black hole accretion but also the formation of a compact $\approx 500 \, \rm pc$ stellar bulge. At this point, the velocity dispersions of the stellar bulges are high $\approx 500 \, \rm km \, s^{-1}$, (ii) as enough AGN energy is coupled to the gas in the nucleus and the mass is expelled in brief AGN-driven outflows, the stellar bulges start to unbind and expand until they reach radii $\approx 5 \, \rm kpc$, (iii) as the gas fractions drop, dry mergers start contributing significantly to size-growth.
Note that another mechanism which can aid in the formation of stellar cores at the centre of massive galaxies is `scouring' by binary black holes \citep{Begelman_1980}, which we can, however, not follow with our simulations due to insufficient resolution.
To some extent, the size growth of massive galaxies may be the result of multiple, independent physical processes operating at different stages in their evolution.

Another interesting implication of our findings relates to the possible upturn in the $M_{\rm BH} \-- \sigma_{\star}$ relation seen for $\sigma_{\star} \gtrsim 270 \, \rm km \, s^{-1}$ \citep[e.g.][]{Kormendy_2013}. 
In our simulations, we find a drop in the peak circular velocity from $\lesssim 500 \, \rm km \, s^{-1}$ to $\lesssim 300 \, \rm km \, s^{-1}$ by $z \approx 3$.
Thus, while the black hole mass doesn't change significantly, the velocity dispersion drops considerably, resulting in an upturn in the $M_{\rm BH} \-- \sigma_{\star}$ relation.

The observed $M_{\rm BH} \-- \sigma_{\star}$ saturates at $\sigma_{\star} \gtrsim 270 \, \rm km \, s^{-1}$ \citep[e.g.][]{Kormendy_2013}, when the morphology of elliptical galaxies switches from predominantly coreless to mainly cored. This finding can be explained by our theoretical results; when black holes grow to sufficiently high masses, they release energies comparable to the binding energy of their host galaxies. As the stellar component expands in response to AGN-driven gas ejection, the massive galaxy develops a cored stellar profile and the stellar velocity dispersion in the central regions drops. AGN feedback is here responsible for both the formation of a core and the drop in the velocity dispersion necessary to result in an upturn in the $M_{\rm BH} \-- \sigma_{\star}$.
One prediction of our scenario is that the characteristic stellar velocity dispersion at which the $M_{\rm BH} \-- \sigma_{\star}$ saturates should evolve with redshift. At $z \gtrsim 5$, the amount of AGN energy released is not sufficient and there has not been enough time to initiate strong galaxy size growth. Thus, at this point, we predict that the $M_{\rm BH} \-- \sigma_{\star}$ should persist even to velocity dispersions $\sigma_{\star} \gtrsim 400 \, \rm km \, s^{-1}$. Since those systems are likely to be the first in which supermassive black holes form and grow most rapidly, they should then be the first to be unbound by AGN feedback. The population of systems with $\sigma_{\star} \gtrsim 400 \, \rm km \, s^{-1}$ should then start thinning out and an upturn in the $M_{\rm BH} \-- \sigma_{\star}$ relation should become visible. Testing the importance of this scenario in simulations following large cosmological boxes will be important to develop more rigorous observational diagnostics.

We have shown that our adopted AGN feedback model is, if anything, too weak. It is thus unlikely that the size growth we find for the quasar host galaxies is driven by too strong AGN feedback.
There are two possible ways, however, in which our AGN feedback model may amplify galaxy size growth.
One is related to limited spatial resolution and the absence of a cold ISM phase from our simulations. As we increase resolution, we may expect the gas distribution to become clumpier, which may possibly permit AGN outflows to escape without coupling with such large volume \citep[e.g.][]{Bourne_2015, Curtis_2016}. Determining whether this is the case will, however, require an accurate treatment of the cold ISM in massive, high redshift galaxies in future simulations.
Another possible way in which galaxy size growth may become less important than seen in our simulations is if AGN feedback couples to halo gas directly rather than to gas in the galactic nucleus.
For instance, if AGN feedback proceeds through jets which pierce through the ISM, but inflate hot bubbles in the diffuse halo, then gas in the galactic nucleus is likely to be less efficiently expelled.
The heating of halo gas may also suppress cooling onto the central galaxy, which in turn may suppress black hole accretion.
Thus, it will be important to establish whether the same AGN-driven galaxy growth is seen in simulations employing different AGN feedback models.

Another uncertainty in this study is the adopted Bondi-Hoyle-Lyttleton formalism to model the black hole accretion. Due to insufficient resolution, we are unable to resolve the accretion onto the black hole self-consistently and use therefore a subgrid model. The parameters of this model (gas density and sound speed) are estimated from scales that are resolved in the simulation. It has been shown that different choices on how these parameters are measured could lead to black hole masses that differ in two orders of magnitude \citep{Curtis_2015, Negri_2017}. In addition, the Bondi-Hoyle-Lyttleton formalism contains assumptions which are unlikely to be true for realistic black hole accretion. The model assumes, for example, spherical symmetry and does not consider the accretion of a turbulent, non-uniform gas flow \citep{Negri_2017}. Alternative models have been developed where, for instance, the angular momentum of the acrreting gas is taken into account \citep{Tremmel_2017} or where the accretion rate is based on the viscous evolution of the accretion disk \citep{Debuhr_2011, Debuhr_2012}. Improving the assumed model is highly desirable to capture AGN feedback more realistically and test whether AGN-driven galaxy growth will still occur.

\begin{figure*}
\includegraphics[width=\textwidth]{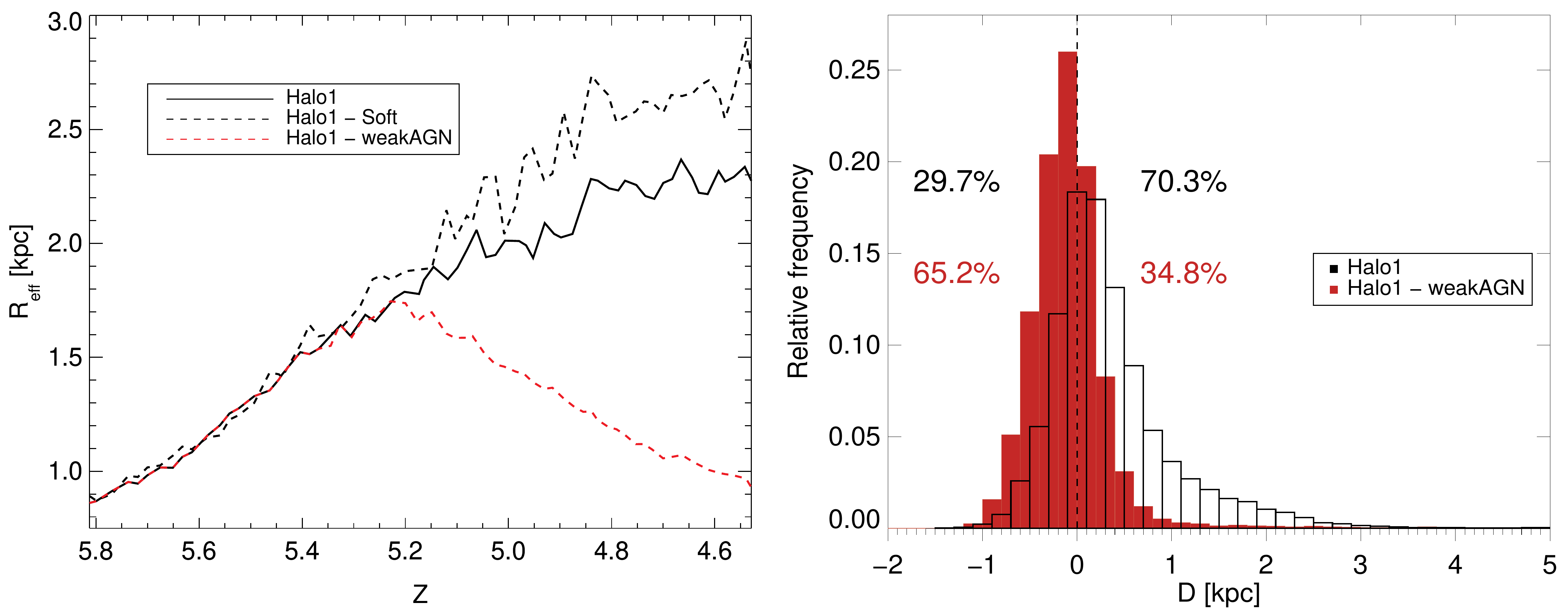}
    \caption{The left-hand panel shows the stellar half mass radius for the simulation with a constant physical softening length (Halo1-Soft), the simulation where AGN feedback is switched off at $z \, = \, 5.2$, (Halo1-weakAGN) and our fiducial simulation Halo1, as a function of redshift. Switching from constant comoving to constant physical softening lengths does not prevent the massive quasar host galaxy from growing in size. The right-hand panel shows the stellar radial distance travelled since $z \,=\, 5.2$ measured for star particles within the central region $r < 1 \, \rm kpc$ at $z \,=\, 5.2$ for Halo1 (black histograms) an d Halo1-weakAGN (red histograms). When AGN feedback is switched off, stellar outward migration ceases, showing that AGN feedback is directly responsible for the outward drift of stars in our simulations. After AGN feedback is switched-off, the tendency is for stars to migrate inwards as gas concentrates in the nucleus, deepening the potential well.}
    \label{fig:r/rsoft}
\end{figure*}

\subsection{Other mechanisms of size growth}
\label{sec:Mechanisms for size growth}
Other potential origins for massive galaxy size growth include physical effects such as major mergers, minor mergers, and adiabatic expansion due the sudden gas expulsion, but there are also potential numerical effects stemming from the non-constant physical size of the gravitational softening length.

In the case of \emph{dry} major mergers, the effective radii and stellar mass both approximately double in the merger remnant ($R_{eff} \propto M_{\star}$). According to this scenario, the high-redshift systems would be uniformly denser than their low-redshift descendants (\citealt{Hernquist_1993, Boylan_2006, Naab_2009}). This kind of growth is not observed for the galaxies in our simulation as can be seen from the right-hand panel in Fig. \ref{fig:size_growth} showing the stellar half mass radius as a function of maximum circular velocity. We see a gradual size growth, that is not expected from a major merger scenario. 
In the minor merger scenario, the compact core remains intact and the overall size growth is then mainly driven by the acquisition of stellar material from less-bound galaxies. \citet{Dubois_2016}, for example, find that the accretion of ex-situ stars drive the galaxy growth at redshift $< 4$.
In our simulations, we, however, find a small fraction of ex-situ stars at least until $z \,=\, 4.5$, when the galaxy has already grown in size significantly, and, in addition, show unambiguously that the compact bulge expands.
Thus, we conclude that the increase in scale radius we find in our simulations is not driven by dry minor mergers.

\citet{Fan_2008} and \citet{Fan_2010} proposed a mechanism in which AGN feedback directly triggers size growth. In this scenario, which is akin to that which may operate via supernova explosions in dwarf galaxies \citep{Navarro_1996}, AGN feedback removes large masses of gas from the central regions on the dynamical timescale or shorter.
The centripetal force, holding orbiting collisionless particles in their orbit, drops substantially and the system relaxes into a new equilibrium configuration with a `puffed-up' stellar and dark matter distributions. Repeating the process accentuates the effect, which allows a significant transformation to be accomplished by recycling a small amount of gas instead of expelling an unfeasibly large amount of gas in one episode. As we have seen in Section \ref{sec:Fluctuating potential}, AGN feedback in the central galaxy in Halo1 removes large amounts of cold gas from the central regions on $\rm Myr$ timescales. 
The final mean size of $\approx 5 \, \rm kpc$ is in line with the expected final radius of $3-5$ kpc after `puffing up' the stellar distribution by feedback processes as described by \cite{Lapi_2018}. 

We explicitly investigate the influence of AGN feedback on the size growth by running the same simulation as described in Section \ref{sec:Evolution Bulge halo1}, but now we switch AGN feedback off at $z \,=\, 5.2$. In Fig. \ref{fig:r/rsoft}, we compare the stellar half mass radius as predicted by {\sc SUBFIND} (see also Section \ref{sec:Evolution circular velocity}) for this simulation with that for the simulation with AGN feedback. We see that the characteristic size of the targeted galaxies in these two simulations become very different at $z < 5.2$, the galaxy where AGN feedback is switched off stops growing in size and gradually decreases in size. The galaxy with AGN feedback has 2.2 times the size of the galaxy with AGN feedback switched off at $z \,=\, 4.6$.

\begin{figure*}
\includegraphics[width=\textwidth]{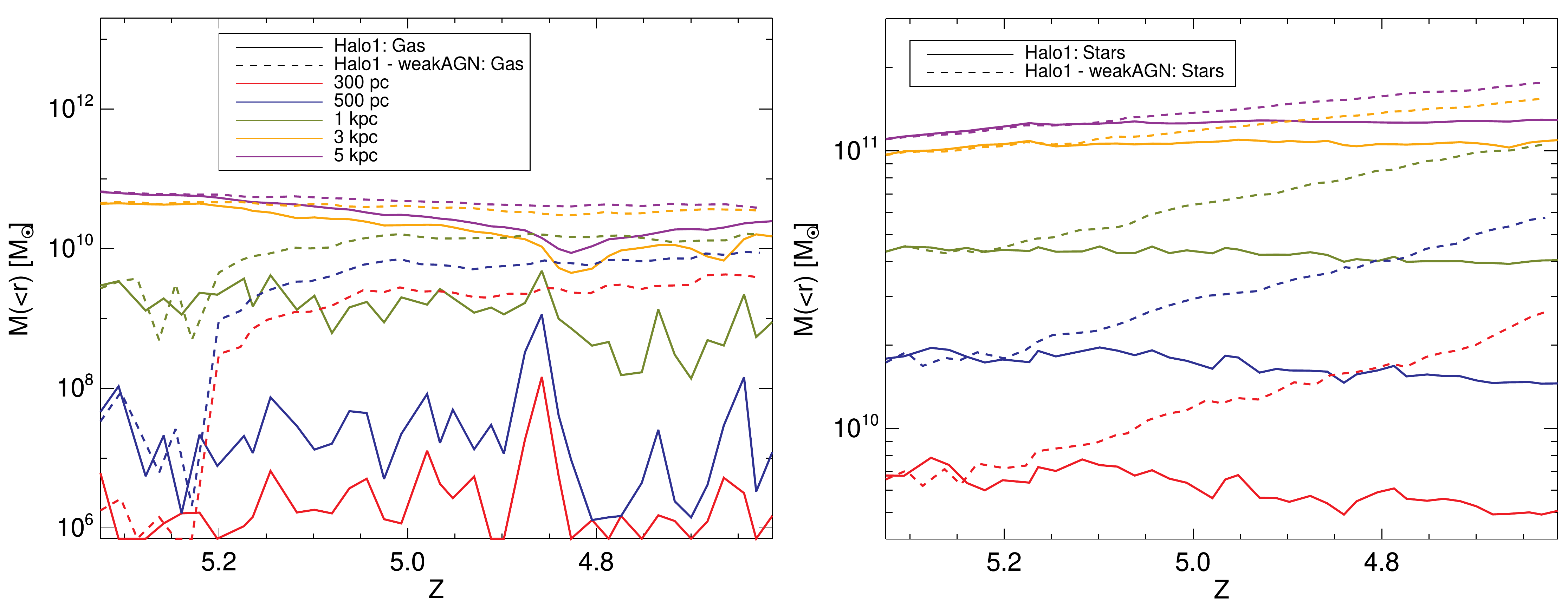}
    \caption{The left- and right-hand panel show the enclosed mass for different radii for the simulation where AGN feedback is switched off at $z \, = \, 5.2$ (dashed lines) and our fiducial simulation (solid lines). The left-hand panel shows the gas component and the right-hand panel shows the stellar component. Clearly, the enclosed gas mass stays highly variable after $z \, \approx \, 5.2$ within $\sim \rm kpc$ scales in the fiducial simulation. The simulation where AGN feedback is switched off stops showing this high variability. The same can be seen in the stellar component although the enclosed stellar mass shows smaller fluctuations. This shows that even the fluctuations within 300 pc (close to the resolution of the simulation) are physically meaningful and driven by AGN driven outflows.
    }
    \label{fig:enclosed_mass_gas_stars_weakAGN}
\end{figure*}

Fig. \ref{fig:enclosed_mass_gas_stars_weakAGN} shows the enclosed gas mass (left-hand panel) and enclosed stellar mass (right-hand panel) at different radial scales, as indicated by the different colours, for our fiducial simulation (Halo1) and the simulation where AGN feedback is switched off at $z \,=\, 5.2$ (Halo1 - weakAGN). We see that the fluctuations in the enclosed gas mass and enclosed stellar mass decrease and almost disappear after AGN feedback is switched off at $z \,=\, 5.2$ for all radii. The fiducial simulation continues to show the fluctuations. In addition, we see that the enclosed stellar mass in the simulation where AGN feedback is switched off gradually increases whereas the enclosed stellar mass gradual decreases in the fiducal simulation (see also Fig. \ref{fig:enclosed_m_2}). This comparison shows that even the fluctuations at a radial scale of 300 pc, close to the resolution of the simulation, are physically meaningful and driven by AGN driven outflows. We also see that the gradual expansion of stellar matter stops when AGN feedback is switched off.

We also check how switching off AGN feedback affects the migration of relic stars. We take the same strategy as described in Section \ref{sec:Ex/in situ stars} and select all stars within the stellar half mass radius at $z \,=\, 5.2$. We then trace these stars down to $z \,=\, 4.6$ and measure the radial distance $D$ they have traveled. The histograms in the right-hand panel of Fig. \ref{fig:r/rsoft} show the stellar radial migration distance for the galaxy with AGN feedback switched off at $z \,=\, 5.2$ (red) and the galaxy with AGN feedback (black). As we have also seen in Fig. \ref{fig:in/ex}, most stars migrate radially outwards in the galaxy with AGN feedback. Since $z \,=\, 5.2$, 70.3\% of the selected stars migrate outwards. The difference with the galaxy with AGN feedback switched off is clear, most stars (65.2\%) migrate inwards and if stars migrate outwards they do not migrate as far as is seen for the galaxy with AGN feedback. AGN feedback thus clearly promotes the outward migration of stars in the galaxy. 

Using the {\sc ILLUSTRIS} simulations, \citet{Wellons_2016} discuss how numerical effects may lead to adiabatic expansion of massive, compact galaxies.
If the softening length is fixed in comoving coordinates, the physical scale over which the gravitational potential is softened increases with time, and stellar particles may migrate outwards as a result. \citet{Wellons_2016} find that such numerical relaxation can change the stellar half mass radius by about 15\% from $z \,=\, 2$ to $z \,=\, 0$, much smaller than the size growth we find in our simulations.

We investigate the influence of numerical relaxation by running the same simulation as described in Section \ref{sec:Evolution Bulge halo1}, but now keeping a constant \emph{physical} softening length at $z < 6$. In Fig. \ref{fig:r/rsoft}, we compare the stellar half mass radius as predicted by {\sc SUBFIND} (see also Section \ref{sec:Evolution circular velocity}) for this simulation with that for the simulation with constant comoving softening length. We see that the characteristic size of the targeted galaxies in these two simulations only becomes systematically different at $z \, < \, 5.5$. The simulation with constant physical softening length results in a galaxy with a greater size as the size found in the simulation with comoving physical softening length. We find that black holes grow more by accretion in the simulation with fixed physical softening length and therefore the injected feedback energy is somewhat larger. This promotes slightly more efficient galaxy
size growth. 
It is clear that there is still significant galaxy growth in the simulation with constant physical softening length. The change in physical softening length that occurs in our fiducial simulations thus contributes only weakly to the galaxy growth we find.
In addition, we see in Fig. \ref{fig:size_growth} that there is almost no size growth in the simulation without AGN feedback, for which numerical relaxation would be the main reason for size growth. The size growth in the galaxy without AGN feedback is about 13\%. We therefore conclude that the softening length cannot explain the observed expansion.

AGN-driven expansion induced by AGN feedback thus seems the main explanation for the observed size growth in our galaxy sample, while other size growth mechanisms may contribute or even dominate at later stages in the evolution of the massive galaxies. 
Also, it is important to stress that we here analyse massive haloes at $z \, = \, 6$, which form in extreme environments which favour rapid black hole growth. The size growth seen might be connected to this rare environment.

Another recent study by \citet{Genel_2018} suggested a similar mechanism for size growth as we found. \citet{Genel_2018} analysed scaling relations and evolution histories of galaxy sizes in Illustris-TNG. The most massive galaxies investigated in \citet{Genel_2018} quench at a stellar mass of $\sim 10^{11} \rm M_{\odot}$. They subsequently grow in size by $\approx 1.5$ dex down to $z \,=\, 0$ as well as in mass, in agreement with the $r \propto M^2$ relation expected from minor dry mergers.

\citet{Genel_2018}, however, find that quenched galaxies that are less massive ($\sim 10^{10.5} \rm M_{\odot}$ at $z \,=\, 0 $) follow a size-mass relation which is significantly steeper than $r \propto M^2$, which may be partly explained by AGN feedback-induced size growth.

\subsection{Cusp/core transformation}
\label{sec:cusp/core}
\begin{figure}
\includegraphics[width=1.0\columnwidth]{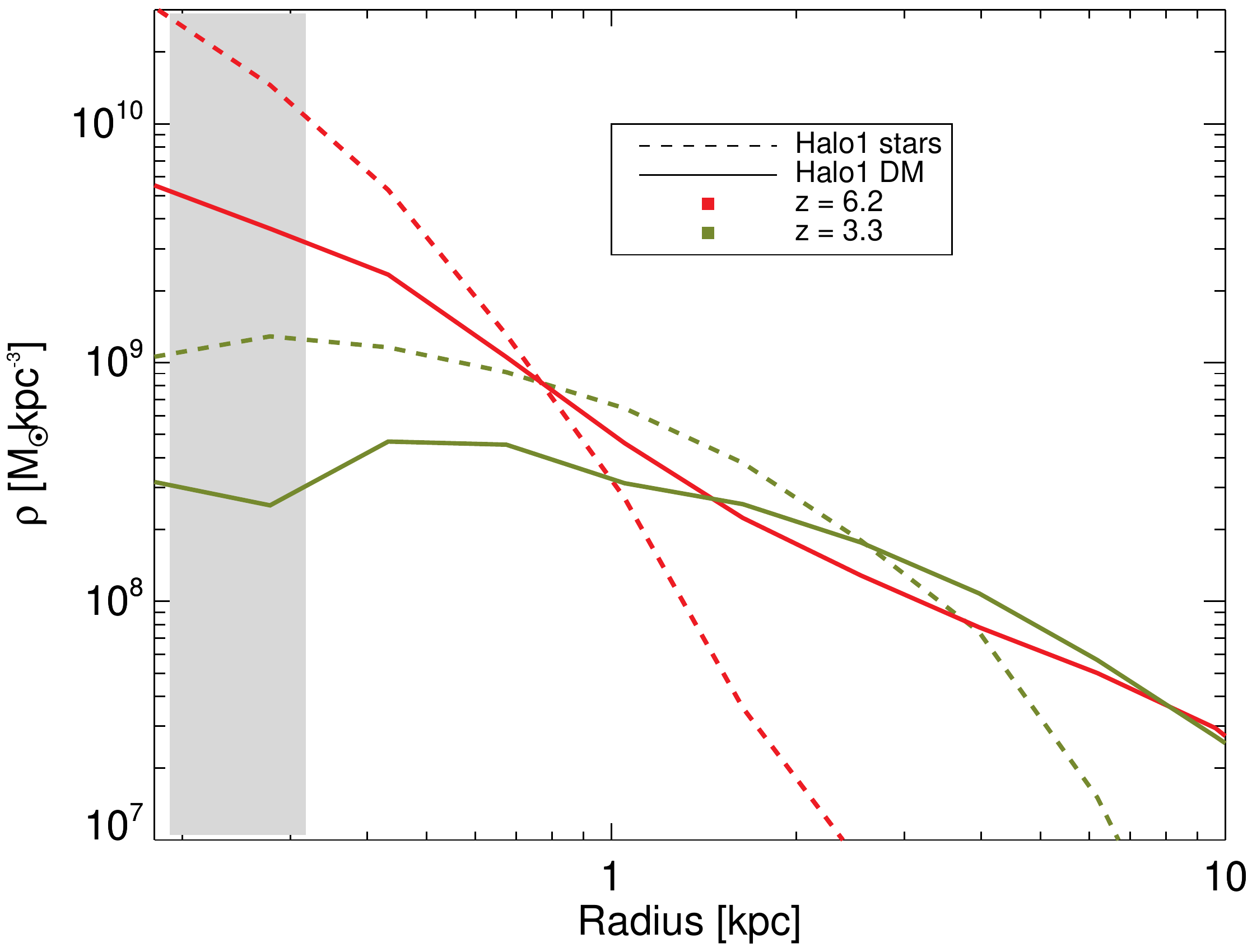}
    \caption{Stellar and dark matter density profiles for Halo1 indicated with dashed and solid lines, respectively. The profiles are shown at $z \, = \, 6.2$ and $z \, = \, 3.3$ in red and green, respectively. The vertical grey region marks the range of the gravitational softening lengths found between $z \, = \, 6.2$ and $z \, = \, 3.3$. The stellar density profile develops a prominent $\sim \, \rm kpc$ sized core by $z \, = \, 3.3$ if AGN feedback is efficient.}
    \label{fig:dens}
\end{figure}

The flattening in density profiles of both the stellar and dark matter component can be seen in Fig. \ref{fig:enclosed_m_2} as a decrease in enclosed mass at small radii (< 1 kpc). Fig. \ref{fig:dens} shows the actual density profiles of stars and dark matter for $z \,=\, 6.2 $ and $z \,=\, 3.3 $ and shows that these profiles flatten significantly over this redshift range. The flattened density profiles are more compatible with the observations of both the stellar and dark matter profiles. Massive elliptical galaxies exhibit very shallow slopes in the stellar surface brightness profiles within small radii of $\approx$ 1 kpc (see \citealt{Quillen_2000, Laine_2003, Graham_2004, Trujillo_2004, Lauer_2005, Cote_2007, Kormendy_2009, Graham_2013}). 

\citet{Pontzen_2012} simulated two dwarf galaxies with a mass of $\sim 10^{10} \rm M_{\odot}$ and investigated their evolution under the influence of supernova feedback. The mass scale of these simulations is different to our scale mass, as our simulations investigate the high mass end of the galaxy mass function. They also found that the potential in the central kiloparsec changes on sub-dynamical time-scales over the redshift interval $4 > z > 2$. These changes were, in contrast to our simulations, induced by feedback from central bursts of star formation. They also show that the fluctuations irreversibly transfer energy into collisionless particles, thus generating the dark matter core, as had been shown much earlier by \citet{Navarro_1996}. Observations that suggest some dwarf galaxies contain AGN in their nuclei \citep[e.g.][]{Pardo_2016, Penny_2018}. It will be interesting to explore whether AGN could shape the stellar and dark matter profiles also in dwarf galaxies in the future.

\citet{Martizzi_2013} also find that AGN feedback can induce a cusp-core transformation, which they showed by using idealised, controlled simulations to follow the response of a massive, cluster-like dark matter halo to feedback from a central black hole of $M_{\rm BH} = 10^{9} \rm M_{\odot}$ over 3.5 Gyr. They fitted a power law to the density profiles, in the region $0.4 < R < 8$ kpc, showing a similar flattening of the slope of the dark matter profile.

\citet{Peirani_2017, Peirani_2018} used the HORIZON and HORIZON-AGN simulations to investigate how AGN feedback affects the evolution of the inner density profiles of massive dark matter haloes ($5 \times 10^{11} \rm M_{\odot}$ - $1 \times 10^{13}\rm M_{\odot}$) and the associated galaxies. They find that the density profiles significantly evolve with time. At $z \approx 1.5$, the mean halo density profiles flatten in the presence of AGN activity. At $z = 0$, the halo (total) density profiles drift back to a more cusped shape, as the AGN feedback efficiency drops. It will be interesting to see if the dark matter profiles in high-z quasar host galaxies experience a similar evolution at lower redshift ($z \leq 2$). 

\section{Summary \& Conclusions}
\label{sec:Conclusions}
We have performed a suite of cosmological, zoom-in hydrodynamical simulations using the moving mesh code AREPO in order to investigate how the stellar component of the galaxies hosting $z \, = \, 6$ quasars evolve after their initial period of rapid black hole growth. Our main findings are:

\begin{itemize}
\item Stellar compact bulges with peak stellar circular velocities in the range $300 \-- 500 \, \rm km \, s^{-1}$ and scale radii of $\approx 300 \, \rm pc$ form by $z \, = \, 6$ even in the presence of strong AGN feedback. While strong outflows are generated \citep[see e.g.][]{Costa_2015}, their impact on the structure of the quasar host galaxy at $z \, = \ 6$ is marginal.
\item The quasar host galaxies remain velocity dispersion dominated ellipticals in the presence of AGN feedback, while they become disc-like in its absence.
\item Despite their high binding energy, all compact bulges in our quasar systems experience considerable size growth from $\approx 300 \, \rm pc$ at $z \, \approx \, 6$ to $\approx 5 \, \rm kpc$ at $z \, \approx \, 3$. 
\item Size growth is accompanied by a drop in the circular velocity, which falls from high values in the range $300 \-- 500 \, \rm km \, s^{-1}$ to $200 \-- 300 \, \rm km \, s^{-1}$. This expansion of the central stellar bulge is seen only if AGN feedback is included.
\item The inner stellar and dark matter density profiles are flattened in the presence of AGN feedback. In the absence of AGN feedback, the compact bulge does not grow significantly in size and the inner dark matter and stellar density profiles instead become steeper.
\item We show that bursty AGN feedback drives fluctuations in the gravitational potential in the innermost few kpc of the quasar host galaxies. We show that these fluctuations lead to the gradual expansion of the collisionless (stellar and dark matter) components.
\item We explicitly verify that switching-off AGN feedback half-way through the simulation leads to an interruption in galaxy size growth and the fluctuations driven by AGN feedback, showing that AGN feedback is directly responsible for the generation of a stellar core.
\end{itemize}

We thus conclude that AGN-driven outflows may be a key ingredient in generating stellar cores at the centre of massive, elliptical galaxies. In fact, the generation of cored stellar profiles is a test to the ability of AGN-driven outflows to remove large amounts of the ISM on short time-scales. This mechanism is also a potential alternative (or may operate in addition) to `scouring' by binary black holes \citep[e.g.][]{Rantala_2018}. AGN-driven outflows may also be responsible for kick-starting the size growth of massive galaxies in the high-redshift Universe, before size growth proceeds through dry mergers. AGN may thus not only suppress the ability of massive galaxies to form stars, but also shape the gravitational potential that facilitates black hole accretion and star formation to begin with. 

A crucial next step will be to investigate the potential of AGN feedback to drive galaxy size growth in cosmological boxes with a large number of galaxies. In addition, higher resolution and additional physical processes (AGN and stellar radiation) as well as an exploration of the effect of different feedback mechanisms will prove essential. 
It is likely that different AGN feedback models will affect galaxy size growth differently, depending on where the energy is effectively injected (halo gas vs. ISM) and on how much gas AGN outflows are able to eject. This means that observational constraints of the inner structure of massive galaxies in particular at $z \gtrsim 2$ are direct tests to the ability of AGN-driven outflows to expel the central ISM and may ultimately be used to narrow down on the coupling efficiency of AGN feedback.

\section*{Acknowledgements}
We thank the anonymous referee for a constructive and helpful report. We thank Benny Trakhtenbrot, Martin Haehnelt, Thorsten Naab, Joop Schaye, Debora Sijacki, Volker Springel and Simon White for helpful discussions and comments on the manuscript.
This work was partially carried out on the Dutch national e-infrastructure with the support of SURF cooperative. We acknowledge that the results of this research have been achieved using the DECI resource Eagle, at PSNC in Poland with support from the PRACE aisbl.

\bibliographystyle{mnras}
\bibliography{ref} 

\begin{thebibliography}{}
\makeatletter
\relax
\def\mn@urlcharsother{\let\do\@makeother \do\$\do\&\do\#\do\^\do\_\do\%\do\~}
\def\mn@doi{\begingroup\mn@urlcharsother \@ifnextchar [ {\mn@doi@}
  {\mn@doi@[]}}
\def\mn@doi@[#1]#2{\def\@tempa{#1}\ifx\@tempa\@empty \href
  {http://dx.doi.org/#2} {doi:#2}\else \href {http://dx.doi.org/#2} {#1}\fi
  \endgroup}
\def\mn@eprint#1#2{\mn@eprint@#1:#2::\@nil}
\def\mn@eprint@arXiv#1{\href {http://arxiv.org/abs/#1} {{\tt arXiv:#1}}}
\def\mn@eprint@dblp#1{\href {http://dblp.uni-trier.de/rec/bibtex/#1.xml}
  {dblp:#1}}
\def\mn@eprint@#1:#2:#3:#4\@nil{\def\@tempa {#1}\def\@tempb {#2}\def\@tempc
  {#3}\ifx \@tempc \@empty \let \@tempc \@tempb \let \@tempb \@tempa \fi \ifx
  \@tempb \@empty \def\@tempb {arXiv}\fi \@ifundefined
  {mn@eprint@\@tempb}{\@tempb:\@tempc}{\expandafter \expandafter \csname
  mn@eprint@\@tempb\endcsname \expandafter{\@tempc}}}

\bibitem[\protect\citeauthoryear{{Ba{\~n}ados} et~al.,}{{Ba{\~n}ados}
  et~al.}{2018}]{Banados_2018}
{Ba{\~n}ados} E.,  et~al., 2018, \mn@doi [\nat] {10.1038/nature25180}, \href
  {http://adsabs.harvard.edu/abs/2018Natur.553..473B} {553, 473}

\bibitem[\protect\citeauthoryear{{Barber}, {Schaye}, {Bower}, {Crain},
  {Schaller}  \& {Theuns}}{{Barber} et~al.}{2016}]{Barber_2016}
{Barber} C.,  {Schaye} J.,  {Bower} R.~G.,  {Crain} R.~A.,  {Schaller} M.,
  {Theuns} T.,  2016, \mn@doi [\mnras] {10.1093/mnras/stw1018}, \href
  {http://adsabs.harvard.edu/abs/2016MNRAS.460.1147B} {460, 1147}

\bibitem[\protect\citeauthoryear{{Begelman}, {Blandford}  \& {Rees}}{{Begelman}
  et~al.}{1980}]{Begelman_1980}
{Begelman} M.~C.,  {Blandford} R.~D.,   {Rees} M.~J.,  1980, \mn@doi [\nat]
  {10.1038/287307a0}, \href {http://adsabs.harvard.edu/abs/1980Natur.287..307B}
  {287, 307}

\bibitem[\protect\citeauthoryear{{Booth} \& {Schaye}}{{Booth} \&
  {Schaye}}{2009}]{Booth_2009}
{Booth} C.~M.,  {Schaye} J.,  2009, \mn@doi [\mnras]
  {10.1111/j.1365-2966.2009.15043.x}, \href
  {http://adsabs.harvard.edu/abs/2009MNRAS.398...53B} {398, 53}

\bibitem[\protect\citeauthoryear{{Bourne}, {Zubovas}  \& {Nayakshin}}{{Bourne}
  et~al.}{2015}]{Bourne_2015}
{Bourne} M.~A.,  {Zubovas} K.,   {Nayakshin} S.,  2015, \mn@doi [\mnras]
  {10.1093/mnras/stv1730}, \href
  {http://adsabs.harvard.edu/abs/2015MNRAS.453.1829B} {453, 1829}

\bibitem[\protect\citeauthoryear{{Boylan-Kolchin}, {Ma}  \&
  {Quataert}}{{Boylan-Kolchin} et~al.}{2006}]{Boylan_2006}
{Boylan-Kolchin} M.,  {Ma} C.-P.,   {Quataert} E.,  2006, \mn@doi [\mnras]
  {10.1111/j.1365-2966.2006.10379.x}, \href
  {http://adsabs.harvard.edu/abs/2006MNRAS.369.1081B} {369, 1081}

\bibitem[\protect\citeauthoryear{{Choi}, {Somerville}, {Ostriker}, {Naab}  \&
  {Hirschmann}}{{Choi} et~al.}{2018}]{Choi_2018}
{Choi} E.,  {Somerville} R.~S.,  {Ostriker} J.~P.,  {Naab} T.,   {Hirschmann}
  M.,  2018, \mn@doi [\apj] {10.3847/1538-4357/aae076}, \href
  {http://adsabs.harvard.edu/abs/2018ApJ...866...91C} {866, 91}

\bibitem[\protect\citeauthoryear{{Correa}, {Schaye}, {Clauwens}, {Bower},
  {Crain}, {Schaller}, {Theuns}  \& {Thob}}{{Correa}
  et~al.}{2017}]{Correa_2017}
{Correa} C.~A.,  {Schaye} J.,  {Clauwens} B.,  {Bower} R.~G.,  {Crain} R.~A.,
  {Schaller} M.,  {Theuns} T.,   {Thob} A.~C.~R.,  2017, \mn@doi [\mnras]
  {10.1093/mnrasl/slx133}, \href
  {http://adsabs.harvard.edu/abs/2017MNRAS.472L..45C} {472, L45}

\bibitem[\protect\citeauthoryear{{Costa}, {Sijacki}, {Trenti}  \&
  {Haehnelt}}{{Costa} et~al.}{2014}]{Costa_2014}
{Costa} T.,  {Sijacki} D.,  {Trenti} M.,   {Haehnelt} M.~G.,  2014, \mn@doi
  [\mnras] {10.1093/mnras/stu101}, \href
  {http://adsabs.harvard.edu/abs/2014MNRAS.439.2146C} {439, 2146}

\bibitem[\protect\citeauthoryear{{Costa}, {Sijacki}  \& {Haehnelt}}{{Costa}
  et~al.}{2015}]{Costa_2015}
{Costa} T.,  {Sijacki} D.,   {Haehnelt} M.~G.,  2015, \mn@doi [\mnras]
  {10.1093/mnrasl/slu193}, \href
  {http://adsabs.harvard.edu/abs/2015MNRAS.448L..30C} {448, L30}

\bibitem[\protect\citeauthoryear{{Costa}, {Rosdahl}, {Sijacki}  \&
  {Haehnelt}}{{Costa} et~al.}{2018}]{Costa_2018}
{Costa} T.,  {Rosdahl} J.,  {Sijacki} D.,   {Haehnelt} M.~G.,  2018, \mn@doi
  [\mnras] {10.1093/mnras/sty1514}, \href
  {http://adsabs.harvard.edu/abs/2018MNRAS.479.2079C} {479, 2079}

\bibitem[\protect\citeauthoryear{{C{\^o}t{\'e}} et~al.,}{{C{\^o}t{\'e}}
  et~al.}{2007}]{Cote_2007}
{C{\^o}t{\'e}} P.,  et~al., 2007, \mn@doi [\apj] {10.1086/522822}, \href
  {http://adsabs.harvard.edu/abs/2007ApJ...671.1456C} {671, 1456}

\bibitem[\protect\citeauthoryear{{Curtis} \& {Sijacki}}{{Curtis} \&
  {Sijacki}}{2015}]{Curtis_2015}
{Curtis} M.,  {Sijacki} D.,  2015, \mn@doi [\mnras] {10.1093/mnras/stv2246},
  \href {https://ui.adsabs.harvard.edu/abs/2015MNRAS.454.3445C} {454, 3445}

\bibitem[\protect\citeauthoryear{{Curtis} \& {Sijacki}}{{Curtis} \&
  {Sijacki}}{2016}]{Curtis_2016}
{Curtis} M.,  {Sijacki} D.,  2016, \mn@doi [\mnras] {10.1093/mnrasl/slv199},
  \href {http://adsabs.harvard.edu/abs/2016MNRAS.457L..34C} {457, L34}

\bibitem[\protect\citeauthoryear{{Dav{\'e}}, {Angl{\'e}s-Alc{\'a}zar},
  {Narayanan}, {Li}, {Rafieferantsoa}  \& {Appleby}}{{Dav{\'e}}
  et~al.}{2019}]{Dave_2019}
{Dav{\'e}} R.,  {Angl{\'e}s-Alc{\'a}zar} D.,  {Narayanan} D.,  {Li} Q.,
  {Rafieferantsoa} M.~H.,   {Appleby} S.,  2019, \mn@doi [\mnras]
  {10.1093/mnras/stz937}, \href
  {https://ui.adsabs.harvard.edu/abs/2019MNRAS.486.2827D} {486, 2827}

\bibitem[\protect\citeauthoryear{{Debuhr}, {Quataert}  \& {Ma}}{{Debuhr}
  et~al.}{2011}]{Debuhr_2011}
{Debuhr} J.,  {Quataert} E.,   {Ma} C.-P.,  2011, \mn@doi [\mnras]
  {10.1111/j.1365-2966.2010.17992.x}, \href
  {https://ui.adsabs.harvard.edu/abs/2011MNRAS.412.1341D} {412, 1341}

\bibitem[\protect\citeauthoryear{{Debuhr}, {Quataert}  \& {Ma}}{{Debuhr}
  et~al.}{2012}]{Debuhr_2012}
{Debuhr} J.,  {Quataert} E.,   {Ma} C.-P.,  2012, \mn@doi [\mnras]
  {10.1111/j.1365-2966.2011.20187.x}, \href
  {https://ui.adsabs.harvard.edu/abs/2012MNRAS.420.2221D} {420, 2221}

\bibitem[\protect\citeauthoryear{{Decarli}, {Falomo}, {Treves}, {Labita},
  {Kotilainen}  \& {Scarpa}}{{Decarli} et~al.}{2010}]{Decarli_2010}
{Decarli} R.,  {Falomo} R.,  {Treves} A.,  {Labita} M.,  {Kotilainen} J.~K.,
  {Scarpa} R.,  2010, \mn@doi [\mnras] {10.1111/j.1365-2966.2009.16049.x},
  \href {http://adsabs.harvard.edu/abs/2010MNRAS.402.2453D} {402, 2453}

\bibitem[\protect\citeauthoryear{{Di Matteo}, {Springel}  \& {Hernquist}}{{Di
  Matteo} et~al.}{2005}]{DiMatteo_2005}
{Di Matteo} T.,  {Springel} V.,   {Hernquist} L.,  2005, \mn@doi [\nat]
  {10.1038/nature03335}, \href
  {http://adsabs.harvard.edu/abs/2005Natur.433..604D} {433, 604}

\bibitem[\protect\citeauthoryear{{Di Matteo}, {Colberg}, {Springel},
  {Hernquist}  \& {Sijacki}}{{Di Matteo} et~al.}{2008}]{DiMatteo_2008}
{Di Matteo} T.,  {Colberg} J.,  {Springel} V.,  {Hernquist} L.,   {Sijacki} D.,
   2008, \mn@doi [\apj] {10.1086/524921}, \href
  {http://adsabs.harvard.edu/abs/2008ApJ...676...33D} {676, 33}

\bibitem[\protect\citeauthoryear{{Di Matteo}, {Khandai}, {DeGraf}, {Feng},
  {Croft}, {Lopez}  \& {Springel}}{{Di Matteo} et~al.}{2012}]{DiMatteo_2012}
{Di Matteo} T.,  {Khandai} N.,  {DeGraf} C.,  {Feng} Y.,  {Croft} R.~A.~C.,
  {Lopez} J.,   {Springel} V.,  2012, \mn@doi [\apjl]
  {10.1088/2041-8205/745/2/L29}, \href
  {http://adsabs.harvard.edu/abs/2012ApJ...745L..29D} {745, L29}

\bibitem[\protect\citeauthoryear{{Di Matteo}, {Croft}, {Feng}, {Waters}  \&
  {Wilkins}}{{Di Matteo} et~al.}{2017}]{DiMatteo_2017}
{Di Matteo} T.,  {Croft} R.~A.~C.,  {Feng} Y.,  {Waters} D.,   {Wilkins} S.,
  2017, \mn@doi [\mnras] {10.1093/mnras/stx319}, \href
  {http://adsabs.harvard.edu/abs/2017MNRAS.467.4243D} {467, 4243}

\bibitem[\protect\citeauthoryear{{Dubois}, {Pichon}, {Haehnelt}, {Kimm},
  {Slyz}, {Devriendt}  \& {Pogosyan}}{{Dubois} et~al.}{2012}]{Dubois_2012}
{Dubois} Y.,  {Pichon} C.,  {Haehnelt} M.,  {Kimm} T.,  {Slyz} A.,  {Devriendt}
  J.,   {Pogosyan} D.,  2012, \mn@doi [\mnras]
  {10.1111/j.1365-2966.2012.21160.x}, \href
  {http://adsabs.harvard.edu/abs/2012MNRAS.423.3616D} {423, 3616}

\bibitem[\protect\citeauthoryear{{Dubois} et~al.,}{{Dubois}
  et~al.}{2014}]{Dubois_2014}
{Dubois} Y.,  et~al., 2014, \mn@doi [\mnras] {10.1093/mnras/stu1227}, \href
  {https://ui.adsabs.harvard.edu/abs/2014MNRAS.444.1453D} {444, 1453}

\bibitem[\protect\citeauthoryear{{Dubois}, {Peirani}, {Pichon}, {Devriendt},
  {Gavazzi}, {Welker}  \& {Volonteri}}{{Dubois} et~al.}{2016}]{Dubois_2016}
{Dubois} Y.,  {Peirani} S.,  {Pichon} C.,  {Devriendt} J.,  {Gavazzi} R.,
  {Welker} C.,   {Volonteri} M.,  2016, \mn@doi [\mnras]
  {10.1093/mnras/stw2265}, \href
  {http://adsabs.harvard.edu/abs/2016MNRAS.463.3948D} {463, 3948}

\bibitem[\protect\citeauthoryear{{Efstathiou} \& {Rees}}{{Efstathiou} \&
  {Rees}}{1988}]{Efstathiou_1988}
{Efstathiou} G.,  {Rees} M.~J.,  1988, \mn@doi [\mnras]
  {10.1093/mnras/230.1.5P}, \href
  {http://adsabs.harvard.edu/abs/1988MNRAS.230P...5E} {230, 5p}

\bibitem[\protect\citeauthoryear{{Fan} et~al.,}{{Fan} et~al.}{2004}]{Fan_2004}
{Fan} X.,  et~al., 2004, \mn@doi [\aj] {10.1086/422434}, \href
  {http://adsabs.harvard.edu/abs/2004AJ....128..515F} {128, 515}

\bibitem[\protect\citeauthoryear{{Fan}, {Lapi}, {De Zotti}  \& {Danese}}{{Fan}
  et~al.}{2008}]{Fan_2008}
{Fan} L.,  {Lapi} A.,  {De Zotti} G.,   {Danese} L.,  2008, \mn@doi [\apjl]
  {10.1086/595784}, \href {http://adsabs.harvard.edu/abs/2008ApJ...689L.101F}
  {689, L101}

\bibitem[\protect\citeauthoryear{{Fan}, {Lapi}, {Bressan}, {Bernardi}, {De
  Zotti}  \& {Danese}}{{Fan} et~al.}{2010}]{Fan_2010}
{Fan} L.,  {Lapi} A.,  {Bressan} A.,  {Bernardi} M.,  {De Zotti} G.,   {Danese}
  L.,  2010, \mn@doi [\apj] {10.1088/0004-637X/718/2/1460}, \href
  {http://adsabs.harvard.edu/abs/2010ApJ...718.1460F} {718, 1460}

\bibitem[\protect\citeauthoryear{{Faucher-Gigu{\`e}re}, {Lidz}, {Zaldarriaga}
  \& {Hernquist}}{{Faucher-Gigu{\`e}re} et~al.}{2009}]{Faucher-Giguere_2009}
{Faucher-Gigu{\`e}re} C.-A.,  {Lidz} A.,  {Zaldarriaga} M.,   {Hernquist} L.,
  2009, \mn@doi [\apj] {10.1088/0004-637X/703/2/1416}, \href
  {http://adsabs.harvard.edu/abs/2009ApJ...703.1416F} {703, 1416}

\bibitem[\protect\citeauthoryear{{Feng}, {Di-Matteo}, {Croft}, {Bird},
  {Battaglia}  \& {Wilkins}}{{Feng} et~al.}{2016}]{Feng_2016}
{Feng} Y.,  {Di-Matteo} T.,  {Croft} R.~A.,  {Bird} S.,  {Battaglia} N.,
  {Wilkins} S.,  2016, \mn@doi [\mnras] {10.1093/mnras/stv2484}, \href
  {https://ui.adsabs.harvard.edu/abs/2016MNRAS.455.2778F} {455, 2778}

\bibitem[\protect\citeauthoryear{{Frigo}, {Naab}, {Hirschmann}, {Choi},
  {Somerville}, {Krajnovic}, {Dav{\'e}}  \& {Cappellari}}{{Frigo}
  et~al.}{2018}]{Frigo_2018}
{Frigo} M.,  {Naab} T.,  {Hirschmann} M.,  {Choi} E.,  {Somerville} R.,
  {Krajnovic} D.,  {Dav{\'e}} R.,   {Cappellari} M.,  2018, \mnras \,
  (submitted), \href {http://adsabs.harvard.edu/abs/2018arXiv181111059F} {}

\bibitem[\protect\citeauthoryear{{Gallerani}, {Fan}, {Maiolino}  \&
  {Pacucci}}{{Gallerani} et~al.}{2017}]{Gallerani_2017}
{Gallerani} S.,  {Fan} X.,  {Maiolino} R.,   {Pacucci} F.,  2017, \mn@doi
  [\pasa] {10.1017/pasa.2017.14}, \href
  {http://adsabs.harvard.edu/abs/2017PASA...34...22G} {34, e022}

\bibitem[\protect\citeauthoryear{{Genel} et~al.,}{{Genel}
  et~al.}{2018}]{Genel_2018}
{Genel} S.,  et~al., 2018, \mn@doi [\mnras] {10.1093/mnras/stx3078}, \href
  {http://adsabs.harvard.edu/abs/2018MNRAS.474.3976G} {474, 3976}

\bibitem[\protect\citeauthoryear{{Graham}}{{Graham}}{2004}]{Graham_2004}
{Graham} A.~W.,  2004, \mn@doi [\apjl] {10.1086/424928}, \href
  {http://adsabs.harvard.edu/abs/2004ApJ...613L..33G} {613, L33}

\bibitem[\protect\citeauthoryear{Graham}{Graham}{2013}]{Graham_2013}
Graham A.~W.,  2013, Elliptical and Disk Galaxy Structure and Modern Scaling
  Laws.
Springer Netherlands, Dordrecht, pp 91--139,
  \mn@doi{10.1007/978-94-007-5609-0_2}, \url
  {https://doi.org/10.1007/978-94-007-5609-0_2}

\bibitem[\protect\citeauthoryear{{Hernquist}, {Spergel}  \& {Heyl}}{{Hernquist}
  et~al.}{1993}]{Hernquist_1993}
{Hernquist} L.,  {Spergel} D.~N.,   {Heyl} J.~S.,  1993, \mn@doi [\apj]
  {10.1086/173247}, \href {http://adsabs.harvard.edu/abs/1993ApJ...416..415H}
  {416, 415}

\bibitem[\protect\citeauthoryear{{Ishibashi}, {Fabian}  \&
  {Canning}}{{Ishibashi} et~al.}{2013}]{Ishibashi_2013}
{Ishibashi} W.,  {Fabian} A.~C.,   {Canning} R.~E.~A.,  2013, \mn@doi [\mnras]
  {10.1093/mnras/stt333}, \href
  {http://adsabs.harvard.edu/abs/2013MNRAS.431.2350I} {431, 2350}

\bibitem[\protect\citeauthoryear{{Johansson}, {Naab}  \&
  {Ostriker}}{{Johansson} et~al.}{2012}]{Johansson_2012}
{Johansson} P.~H.,  {Naab} T.,   {Ostriker} J.~P.,  2012, \mn@doi [\apj]
  {10.1088/0004-637X/754/2/115}, \href
  {http://adsabs.harvard.edu/abs/2012ApJ...754..115J} {754, 115}

\bibitem[\protect\citeauthoryear{{Katz}, {Weinberg}  \& {Hernquist}}{{Katz}
  et~al.}{1996}]{Katz_1996}
{Katz} N.,  {Weinberg} D.~H.,   {Hernquist} L.,  1996, \mn@doi [\apjs]
  {10.1086/192305}, \href {http://adsabs.harvard.edu/abs/1996ApJS..105...19K}
  {105, 19}

\bibitem[\protect\citeauthoryear{{Kormendy} \& {Ho}}{{Kormendy} \&
  {Ho}}{2013}]{Kormendy_2013}
{Kormendy} J.,  {Ho} L.~C.,  2013, \mn@doi [\araa]
  {10.1146/annurev-astro-082708-101811}, \href
  {http://adsabs.harvard.edu/abs/2013ARA%26A..51..511K} {51, 511}

\bibitem[\protect\citeauthoryear{{Kormendy}, {Fisher}, {Cornell}  \&
  {Bender}}{{Kormendy} et~al.}{2009}]{Kormendy_2009}
{Kormendy} J.,  {Fisher} D.~B.,  {Cornell} M.~E.,   {Bender} R.,  2009, \mn@doi
  [\apjs] {10.1088/0067-0049/182/1/216}, \href
  {http://adsabs.harvard.edu/abs/2009ApJS..182..216K} {182, 216}

\bibitem[\protect\citeauthoryear{{Laine}, {van der Marel}, {Lauer}, {Postman},
  {O'Dea}  \& {Owen}}{{Laine} et~al.}{2003}]{Laine_2003}
{Laine} S.,  {van der Marel} R.~P.,  {Lauer} T.~R.,  {Postman} M.,  {O'Dea}
  C.~P.,   {Owen} F.~N.,  2003, \mn@doi [\aj] {10.1086/345823}, \href
  {http://adsabs.harvard.edu/abs/2003AJ....125..478L} {125, 478}

\bibitem[\protect\citeauthoryear{{Lapi} et~al.,}{{Lapi}
  et~al.}{2018}]{Lapi_2018}
{Lapi} A.,  et~al., 2018, \mn@doi [\apj] {10.3847/1538-4357/aab6af}, \href
  {http://adsabs.harvard.edu/abs/2018ApJ...857...22L} {857, 22}

\bibitem[\protect\citeauthoryear{{Lauer} et~al.,}{{Lauer}
  et~al.}{2005}]{Lauer_2005}
{Lauer} T.~R.,  et~al., 2005, \mn@doi [\aj] {10.1086/429565}, \href
  {http://adsabs.harvard.edu/abs/2005AJ....129.2138L} {129, 2138}

\bibitem[\protect\citeauthoryear{{Martizzi}, {Teyssier}  \& {Moore}}{{Martizzi}
  et~al.}{2013}]{Martizzi_2013}
{Martizzi} D.,  {Teyssier} R.,   {Moore} B.,  2013, \mn@doi [\mnras]
  {10.1093/mnras/stt297}, \href
  {http://adsabs.harvard.edu/abs/2013MNRAS.432.1947M} {432, 1947}

\bibitem[\protect\citeauthoryear{{McConnell} \& {Ma}}{{McConnell} \&
  {Ma}}{2013}]{McConnell_2013}
{McConnell} N.~J.,  {Ma} C.-P.,  2013, \mn@doi [\apj]
  {10.1088/0004-637X/764/2/184}, \href
  {http://adsabs.harvard.edu/abs/2013ApJ...764..184M} {764, 184}

\bibitem[\protect\citeauthoryear{{Merloni} et~al.,}{{Merloni}
  et~al.}{2010}]{Merloni_2010}
{Merloni} A.,  et~al., 2010, \mn@doi [\apj] {10.1088/0004-637X/708/1/137},
  \href {http://adsabs.harvard.edu/abs/2010ApJ...708..137M} {708, 137}

\bibitem[\protect\citeauthoryear{{Mortlock} et~al.,}{{Mortlock}
  et~al.}{2011}]{Mortlock_2011}
{Mortlock} D.~J.,  et~al., 2011, \mn@doi [\nat] {10.1038/nature10159}, \href
  {http://adsabs.harvard.edu/abs/2011Natur.474..616M} {474, 616}

\bibitem[\protect\citeauthoryear{{Moster}, {Naab}  \& {White}}{{Moster}
  et~al.}{2013}]{Moster_2013}
{Moster} B.~P.,  {Naab} T.,   {White} S.~D.~M.,  2013, \mn@doi [\mnras]
  {10.1093/mnras/sts261}, \href
  {http://adsabs.harvard.edu/abs/2013MNRAS.428.3121M} {428, 3121}

\bibitem[\protect\citeauthoryear{{Naab}, {Johansson}  \& {Ostriker}}{{Naab}
  et~al.}{2009}]{Naab_2009}
{Naab} T.,  {Johansson} P.~H.,   {Ostriker} J.~P.,  2009, \mn@doi [\apjl]
  {10.1088/0004-637X/699/2/L178}, \href
  {http://adsabs.harvard.edu/abs/2009ApJ...699L.178N} {699, L178}

\bibitem[\protect\citeauthoryear{{Navarro}, {Eke}  \& {Frenk}}{{Navarro}
  et~al.}{1996}]{Navarro_1996}
{Navarro} J.~F.,  {Eke} V.~R.,   {Frenk} C.~S.,  1996, \mn@doi [\mnras]
  {10.1093/mnras/283.3.L72}, \href
  {http://adsabs.harvard.edu/abs/1996MNRAS.283L..72N} {283, L72}

\bibitem[\protect\citeauthoryear{{Negri} \& {Volonteri}}{{Negri} \&
  {Volonteri}}{2017}]{Negri_2017}
{Negri} A.,  {Volonteri} M.,  2017, \mn@doi [\mnras] {10.1093/mnras/stx362},
  \href {https://ui.adsabs.harvard.edu/abs/2017MNRAS.467.3475N} {467, 3475}

\bibitem[\protect\citeauthoryear{{Ogiya} \& {Mori}}{{Ogiya} \&
  {Mori}}{2014}]{Ogiya_2014}
{Ogiya} G.,  {Mori} M.,  2014, \mn@doi [\apj] {10.1088/0004-637X/793/1/46},
  \href {http://adsabs.harvard.edu/abs/2014ApJ...793...46O} {793, 46}

\bibitem[\protect\citeauthoryear{{Oser}, {Ostriker}, {Naab}, {Johansson}  \&
  {Burkert}}{{Oser} et~al.}{2010}]{Oser_2010}
{Oser} L.,  {Ostriker} J.~P.,  {Naab} T.,  {Johansson} P.~H.,   {Burkert} A.,
  2010, \mn@doi [\apj] {10.1088/0004-637X/725/2/2312}, \href
  {http://adsabs.harvard.edu/abs/2010ApJ...725.2312O} {725, 2312}

\bibitem[\protect\citeauthoryear{{Pardo} et~al.,}{{Pardo}
  et~al.}{2016}]{Pardo_2016}
{Pardo} K.,  et~al., 2016, \mn@doi [\apj] {10.3847/0004-637X/831/2/203}, \href
  {http://adsabs.harvard.edu/abs/2016ApJ...831..203P} {831, 203}

\bibitem[\protect\citeauthoryear{{Peirani} et~al.,}{{Peirani}
  et~al.}{2017}]{Peirani_2017}
{Peirani} S.,  et~al., 2017, \mn@doi [\mnras] {10.1093/mnras/stx2099}, \href
  {http://adsabs.harvard.edu/abs/2017MNRAS.472.2153P} {472, 2153}

\bibitem[\protect\citeauthoryear{{Peirani} et~al.,}{{Peirani}
  et~al.}{2019}]{Peirani_2018}
{Peirani} S.,  et~al., 2019, \mn@doi [\mnras] {10.1093/mnras/sty3475}, \href
  {http://adsabs.harvard.edu/abs/2019MNRAS.483.4615P} {483, 4615}

\bibitem[\protect\citeauthoryear{{Penny} et~al.,}{{Penny}
  et~al.}{2018}]{Penny_2018}
{Penny} S.~J.,  et~al., 2018, \mn@doi [\mnras] {10.1093/mnras/sty202}, \href
  {http://adsabs.harvard.edu/abs/2018MNRAS.476..979P} {476, 979}

\bibitem[\protect\citeauthoryear{{Penoyre}, {Moster}, {Sijacki}  \&
  {Genel}}{{Penoyre} et~al.}{2017}]{Penoyre_2017}
{Penoyre} Z.,  {Moster} B.~P.,  {Sijacki} D.,   {Genel} S.,  2017, \mn@doi
  [\mnras] {10.1093/mnras/stx762}, \href
  {http://adsabs.harvard.edu/abs/2017MNRAS.468.3883P} {468, 3883}

\bibitem[\protect\citeauthoryear{{Planck Collaboration} et~al.,}{{Planck
  Collaboration} et~al.}{2014}]{PlanckCollaboration2014}
{Planck Collaboration} et~al., 2014, \mn@doi [\aap]
  {10.1051/0004-6361/201321591}, \href
  {http://adsabs.harvard.edu/abs/2014A%26A...571A..16P} {571, A16}

\bibitem[\protect\citeauthoryear{{Pontzen} \& {Governato}}{{Pontzen} \&
  {Governato}}{2012}]{Pontzen_2012}
{Pontzen} A.,  {Governato} F.,  2012, \mn@doi [\mnras]
  {10.1111/j.1365-2966.2012.20571.x}, \href
  {http://adsabs.harvard.edu/abs/2012MNRAS.421.3464P} {421, 3464}

\bibitem[\protect\citeauthoryear{{Pontzen} \& {Governato}}{{Pontzen} \&
  {Governato}}{2014}]{Pontzen_2014}
{Pontzen} A.,  {Governato} F.,  2014, \mn@doi [\nat] {10.1038/nature12953},
  \href {http://adsabs.harvard.edu/abs/2014Natur.506..171P} {506, 171}

\bibitem[\protect\citeauthoryear{{Quillen}, {Bower}  \&
  {Stritzinger}}{{Quillen} et~al.}{2000}]{Quillen_2000}
{Quillen} A.~C.,  {Bower} G.~A.,   {Stritzinger} M.,  2000, \mn@doi [\apjs]
  {10.1086/313374}, \href {http://adsabs.harvard.edu/abs/2000ApJS..128...85Q}
  {128, 85}

\bibitem[\protect\citeauthoryear{{Ragone-Figueroa} \&
  {Granato}}{{Ragone-Figueroa} \& {Granato}}{2011}]{RagoneFigueroa_2011}
{Ragone-Figueroa} C.,  {Granato} G.~L.,  2011, \mn@doi [\mnras]
  {10.1111/j.1365-2966.2011.18670.x}, \href
  {http://adsabs.harvard.edu/abs/2011MNRAS.414.3690R} {414, 3690}

\bibitem[\protect\citeauthoryear{{Rantala}, {Johansson}, {Naab}, {Thomas}  \&
  {Frigo}}{{Rantala} et~al.}{2018}]{Rantala_2018}
{Rantala} A.,  {Johansson} P.~H.,  {Naab} T.,  {Thomas} J.,   {Frigo} M.,
  2018, \apjl \, (submitted), \href
  {http://adsabs.harvard.edu/abs/2018arXiv181202732R} {}

\bibitem[\protect\citeauthoryear{{Reines} \& {Volonteri}}{{Reines} \&
  {Volonteri}}{2015}]{Reines_2015}
{Reines} A.~E.,  {Volonteri} M.,  2015, \mn@doi [\apj]
  {10.1088/0004-637X/813/2/82}, \href
  {http://adsabs.harvard.edu/abs/2015ApJ...813...82R} {813, 82}

\bibitem[\protect\citeauthoryear{{Saglia} et~al.,}{{Saglia}
  et~al.}{2016}]{Saglia_2016}
{Saglia} R.~P.,  et~al., 2016, \mn@doi [\apj] {10.3847/0004-637X/818/1/47},
  \href {http://adsabs.harvard.edu/abs/2016ApJ...818...47S} {818, 47}

\bibitem[\protect\citeauthoryear{{Schaye} et~al.,}{{Schaye}
  et~al.}{2015}]{Schaye_2015}
{Schaye} J.,  et~al., 2015, \mn@doi [\mnras] {10.1093/mnras/stu2058}, \href
  {https://ui.adsabs.harvard.edu/abs/2015MNRAS.446..521S} {446, 521}

\bibitem[\protect\citeauthoryear{{Seth} et~al.,}{{Seth}
  et~al.}{2014}]{Seth_2014}
{Seth} A.~C.,  et~al., 2014, \mn@doi [\nat] {10.1038/nature13762}, \href
  {http://adsabs.harvard.edu/abs/2014Natur.513..398S} {513, 398}

\bibitem[\protect\citeauthoryear{{Sijacki}, {Springel}, {Di Matteo}  \&
  {Hernquist}}{{Sijacki} et~al.}{2007}]{Sijacki_2007}
{Sijacki} D.,  {Springel} V.,  {Di Matteo} T.,   {Hernquist} L.,  2007, \mn@doi
  [\mnras] {10.1111/j.1365-2966.2007.12153.x}, \href
  {http://adsabs.harvard.edu/abs/2007MNRAS.380..877S} {380, 877}

\bibitem[\protect\citeauthoryear{{Sijacki}, {Springel}  \&
  {Haehnelt}}{{Sijacki} et~al.}{2009}]{Sijacki_2009}
{Sijacki} D.,  {Springel} V.,   {Haehnelt} M.~G.,  2009, \mn@doi [\mnras]
  {10.1111/j.1365-2966.2009.15452.x}, \href
  {http://adsabs.harvard.edu/abs/2009MNRAS.400..100S} {400, 100}

\bibitem[\protect\citeauthoryear{{Sijacki}, {Vogelsberger}, {Genel},
  {Springel}, {Torrey}, {Snyder}, {Nelson}  \& {Hernquist}}{{Sijacki}
  et~al.}{2015}]{Sijacki_2015}
{Sijacki} D.,  {Vogelsberger} M.,  {Genel} S.,  {Springel} V.,  {Torrey} P.,
  {Snyder} G.~F.,  {Nelson} D.,   {Hernquist} L.,  2015, \mn@doi [\mnras]
  {10.1093/mnras/stv1340}, \href
  {http://adsabs.harvard.edu/abs/2015MNRAS.452..575S} {452, 575}

\bibitem[\protect\citeauthoryear{{Springel}}{{Springel}}{2010}]{Springel_2010}
{Springel} V.,  2010, \mn@doi [\mnras] {10.1111/j.1365-2966.2009.15715.x},
  \href {http://adsabs.harvard.edu/abs/2010MNRAS.401..791S} {401, 791}

\bibitem[\protect\citeauthoryear{{Springel} \& {Hernquist}}{{Springel} \&
  {Hernquist}}{2003}]{Springel_2003}
{Springel} V.,  {Hernquist} L.,  2003, \mn@doi [\mnras]
  {10.1046/j.1365-8711.2003.06206.x}, \href
  {http://adsabs.harvard.edu/abs/2003MNRAS.339..289S} {339, 289}

\bibitem[\protect\citeauthoryear{{Springel}, {White}, {Tormen}  \&
  {Kauffmann}}{{Springel} et~al.}{2001}]{Springel_2001}
{Springel} V.,  {White} S.~D.~M.,  {Tormen} G.,   {Kauffmann} G.,  2001,
  \mn@doi [\mnras] {10.1046/j.1365-8711.2001.04912.x}, \href
  {http://adsabs.harvard.edu/abs/2001MNRAS.328..726S} {328, 726}

\bibitem[\protect\citeauthoryear{{Springel} et~al.,}{{Springel}
  et~al.}{2005}]{Springel_2005}
{Springel} V.,  et~al., 2005, \mn@doi [\nat] {10.1038/nature03597}, \href
  {http://adsabs.harvard.edu/abs/2005Natur.435..629S} {435, 629}

\bibitem[\protect\citeauthoryear{{Teyssier}, {Pontzen}, {Dubois}  \&
  {Read}}{{Teyssier} et~al.}{2013}]{Teyssier_2013}
{Teyssier} R.,  {Pontzen} A.,  {Dubois} Y.,   {Read} J.~I.,  2013, \mn@doi
  [\mnras] {10.1093/mnras/sts563}, \href
  {http://adsabs.harvard.edu/abs/2013MNRAS.429.3068T} {429, 3068}

\bibitem[\protect\citeauthoryear{{Trakhtenbrot} et~al.,}{{Trakhtenbrot}
  et~al.}{2015}]{Trakhtenbrot_2015}
{Trakhtenbrot} B.,  et~al., 2015, \mn@doi [Science] {10.1126/science.aaa4506},
  \href {http://adsabs.harvard.edu/abs/2015Sci...349..168T} {349, 168}

\bibitem[\protect\citeauthoryear{{Tremmel}, {Karcher}, {Governato},
  {Volonteri}, {Quinn}, {Pontzen}, {Anderson}  \& {Bellovary}}{{Tremmel}
  et~al.}{2017}]{Tremmel_2017}
{Tremmel} M.,  {Karcher} M.,  {Governato} F.,  {Volonteri} M.,  {Quinn} T.~R.,
  {Pontzen} A.,  {Anderson} L.,   {Bellovary} J.,  2017, \mn@doi [\mnras]
  {10.1093/mnras/stx1160}, \href
  {https://ui.adsabs.harvard.edu/abs/2017MNRAS.470.1121T} {470, 1121}

\bibitem[\protect\citeauthoryear{{Trujillo}, {Erwin}, {Asensio Ramos}  \&
  {Graham}}{{Trujillo} et~al.}{2004}]{Trujillo_2004}
{Trujillo} I.,  {Erwin} P.,  {Asensio Ramos} A.,   {Graham} A.~W.,  2004,
  \mn@doi [\aj] {10.1086/382712}, \href
  {http://adsabs.harvard.edu/abs/2004AJ....127.1917T} {127, 1917}

\bibitem[\protect\citeauthoryear{{Venemans} et~al.,}{{Venemans}
  et~al.}{2013}]{Venemans_2013}
{Venemans} B.~P.,  et~al., 2013, \mn@doi [\apj] {10.1088/0004-637X/779/1/24},
  \href {http://adsabs.harvard.edu/abs/2013ApJ...779...24V} {779, 24}

\bibitem[\protect\citeauthoryear{{Vogelsberger}, {Genel}, {Sijacki}, {Torrey},
  {Springel}  \& {Hernquist}}{{Vogelsberger} et~al.}{2013}]{Vogelsberger_2013}
{Vogelsberger} M.,  {Genel} S.,  {Sijacki} D.,  {Torrey} P.,  {Springel} V.,
  {Hernquist} L.,  2013, \mn@doi [\mnras] {10.1093/mnras/stt1789}, \href
  {http://adsabs.harvard.edu/abs/2013MNRAS.436.3031V} {436, 3031}

\bibitem[\protect\citeauthoryear{{Volonteri} \& {Rees}}{{Volonteri} \&
  {Rees}}{2006}]{Volonteri_2006}
{Volonteri} M.,  {Rees} M.~J.,  2006, \mn@doi [\apj] {10.1086/507444}, \href
  {http://adsabs.harvard.edu/abs/2006ApJ...650..669V} {650, 669}

\bibitem[\protect\citeauthoryear{{Volonteri} \& {Stark}}{{Volonteri} \&
  {Stark}}{2011}]{Volonteri_2011}
{Volonteri} M.,  {Stark} D.~P.,  2011, \mn@doi [\mnras]
  {10.1111/j.1365-2966.2011.19391.x}, \href
  {http://adsabs.harvard.edu/abs/2011MNRAS.417.2085V} {417, 2085}

\bibitem[\protect\citeauthoryear{{Walsh}, {van den Bosch}, {Gebhardt},
  {Yildirim}, {G{\"u}ltekin}, {Husemann}  \& {Richstone}}{{Walsh}
  et~al.}{2015}]{Walsh_2015}
{Walsh} J.~L.,  {van den Bosch} R.~C.~E.,  {Gebhardt} K.,  {Yildirim} A.,
  {G{\"u}ltekin} K.,  {Husemann} B.,   {Richstone} D.~O.,  2015, \mn@doi [\apj]
  {10.1088/0004-637X/808/2/183}, \href
  {http://adsabs.harvard.edu/abs/2015ApJ...808..183W} {808, 183}

\bibitem[\protect\citeauthoryear{{Walsh}, {van den Bosch}, {Gebhardt},
  {Y{\i}ld{\i}r{\i}m}, {Richstone}, {G{\"u}ltekin}  \& {Husemann}}{{Walsh}
  et~al.}{2016}]{Walsh_2016}
{Walsh} J.~L.,  {van den Bosch} R.~C.~E.,  {Gebhardt} K.,  {Y{\i}ld{\i}r{\i}m}
  A.,  {Richstone} D.~O.,  {G{\"u}ltekin} K.,   {Husemann} B.,  2016, \mn@doi
  [\apj] {10.3847/0004-637X/817/1/2}, \href
  {http://adsabs.harvard.edu/abs/2016ApJ...817....2W} {817, 2}

\bibitem[\protect\citeauthoryear{{Weinberger} et~al.,}{{Weinberger}
  et~al.}{2017}]{Weinberger_2017}
{Weinberger} R.,  et~al., 2017, \mn@doi [\mnras] {10.1093/mnras/stw2944}, \href
  {http://adsabs.harvard.edu/abs/2017MNRAS.465.3291W} {465, 3291}

\bibitem[\protect\citeauthoryear{{Weinberger} et~al.,}{{Weinberger}
  et~al.}{2018}]{Weinberger_2018}
{Weinberger} R.,  et~al., 2018, \mn@doi [\mnras] {10.1093/mnras/sty1733}, \href
  {http://adsabs.harvard.edu/abs/2018MNRAS.479.4056W} {479, 4056}

\bibitem[\protect\citeauthoryear{{Wellons} et~al.,}{{Wellons}
  et~al.}{2016}]{Wellons_2016}
{Wellons} S.,  et~al., 2016, \mn@doi [\mnras] {10.1093/mnras/stv2738}, \href
  {http://adsabs.harvard.edu/abs/2016MNRAS.456.1030W} {456, 1030}

\bibitem[\protect\citeauthoryear{{Willott} et~al.,}{{Willott}
  et~al.}{2010}]{Willott_2010}
{Willott} C.~J.,  et~al., 2010, \mn@doi [\aj] {10.1088/0004-6256/139/3/906},
  \href {http://adsabs.harvard.edu/abs/2010AJ....139..906W} {139, 906}

\bibitem[\protect\citeauthoryear{{Wu} et~al.,}{{Wu} et~al.}{2015}]{Wu_2015}
{Wu} X.-B.,  et~al., 2015, \mn@doi [\nat] {10.1038/nature14241}, \href
  {http://adsabs.harvard.edu/abs/2015Natur.518..512W} {518, 512}

\bibitem[\protect\citeauthoryear{{van Loon} \& {Sansom}}{{van Loon} \&
  {Sansom}}{2015}]{vanLoon_2015}
{van Loon} J.~T.,  {Sansom} A.~E.,  2015, \mn@doi [\mnras]
  {10.1093/mnras/stv1787}, \href
  {http://adsabs.harvard.edu/abs/2015MNRAS.453.2341V} {453, 2341}

\makeatother
\end{thebibliography}


\bsp	
\label{lastpage}
\end{document}